\documentclass[12pt]{iopart}
\usepackage{iopams}  
\usepackage{graphicx}
\usepackage{color}
\usepackage{caption}
\usepackage{subcaption}
\usepackage[dvipsnames]{xcolor}
\usepackage{psfrag,epsfig}
\usepackage[percent]{overpic}
\usepackage{float}
\usepackage{color}
\usepackage{multirow}
\usepackage{tikz}
\usetikzlibrary{arrows.meta,bending,positioning,intersections}
\usepackage{pgfplots}
\usepackage[all]{xy}
\def \fo {{f_{1}}}
\def \ftt {{f_{2,2}}}

\begin{document}

\title[Flux-conserving directed percolation]{Flux-conserving directed percolation}

\author{Barto Cucurull$^{1}$, Greg Huber$^{2}$, Kyle Kawagoe$^{3}$, Marc Pradas$^{1}$, 
Alain Pumir$^{4}$ and Michael Wilkinson$^{1}$}

\address{
$^1$ Department of Mathematics and Statistics,
The Open University, Walton Hall, Milton Keynes, MK7 6AA, England,\\
 $^2$ Chan Zuckerberg Biohub -- San Francisco, 499 Illinois Street, San Francisco, CA 94158, USA,\\
$^3$ Departments of Physics and Mathematics, 
The Ohio State University, Columbus, OH 43210, USA\\
$^4$ Laboratoire de Physique,
 Ecole Normale Sup\'erieure de Lyon, CNRS, Universit\'e de Lyon,
 F-69007, Lyon, France,\\
}
\vspace{10pt}
\ead{
gerghuber@gmail.com,
kawagoe.1@osu.edu,
marc.pradas@open.ac.uk,
alain.pumir@ens-lyon.fr,
m.wilkinson@open.ac.uk
}

\begin{indented}
\item June 2023
\end{indented}

\begin{abstract}
We discuss a model for directed percolation in which the 
flux of material along each bond is a dynamical variable. 
The model includes a physically significant limiting case 
where the total flux of material is conserved. We show that the 
distribution of fluxes is asymptotic to a power law at small fluxes. 
We give an implicit equation for the exponent, in terms of probabilities 
characterising site occupations. In one dimension the site occupations 
are exactly independent, and the model is exactly solvable. In two 
dimensions, the independent-occupation assumption gives a good approximation. 
We explore the relationship between this model and traditional 
models for directed percolation.
\end{abstract}

%\submitto{\jpa}
\maketitle

\section{Introduction}
\label{sec: 1}

Percolation problems were introduced by Broadbent and Hammersley 
in 1957 \cite{Bro+57}. Their paper motivated the study of percolation \cite{Sta+92} by a discussion 
of fluids passing through a disordered medium, such as water penetrating 
through limestone. Forced flow of a liquid through a porous 
medium is central to many interesting and technologically important 
processes involving elution from, or absorption by, random media, 
such as leaching of salts from soil, extraction of oil or gas 
from reservoirs, brewing coffee, or the operation of chromatography 
columns, which stimulated the study of \emph{directed percolation}.
There is a vast literature treating the standard models of directed 
percolation, reviewed in \cite{Hin00}, and
\cite{Gra+79,Jan81,Gra81,Aro+83,Red83,Dom+84,Kin85,Zif+86,
Bax+88,Tak+92,Hub+95,Gra95,Tre+95}
are indicative of the breadth of different approaches to directed percolation, 
and of its wide range of applications. 
The standard percolation models (such as 
bond percolation on a lattice) 
do not take into account 
the mass conservation of a flowing liquid. In this paper we consider 
a generalisation of directed percolation, which includes the flux in a bond 
as a dynamical variable. If the fluxes are ignored, and only the occupancy 
of bonds is considered, then the standard directed percolation model occurs 
as a particular special case. Our generalised model includes a \emph{flux-conserving} 
case which can describe the forced flow of a liquid through a random medium, and 
we shall consider this in some detail. We demonstrate that aspects of the subset of models 
which represents elution by a flux-conserving fluid are exactly solvable in one dimension, 
extending some results obtained (in another context) in \cite{Kaw+17}. Our generalised model
has some similarity to the Scheidegger model for the distribution of 
river catchment basins \cite{Sch67,Hub91}, which will be discussed
in the conclusions. Another, more distantly related, class of models which 
quantify directed transport in random media 
is described in \cite{Nar+93,Wat+96}.

The percolation model introduced by Broadbent and Hammersley uses the 
idea of \lq wetted' bonds. Here we extend this binary notion of wetting by 
modelling the flux $\phi$ of liquid through each bond 
and the probability distribution of such fluxes. 
We argue that, for a quite general class of flux-conserving process 
which involve a forced 
flow through a disordered medium, the distribution of the fluxes has 
some universal characteristics. After penetrating a sufficient distance into 
the network, the probability density $P(\phi)$ for the flux $\phi$ in a channel 
may approach a stationary distribution, with a 
power-law form at small values of $\phi$:
\begin{equation}
\label{eq: 1.1}
P(\phi)\sim \phi^{-\alpha}
\ .
\end{equation}
The existence of power laws is usually associated with critical phenomena, however 
the power-law distribution described by equation (\ref{eq: 1.1}) is a robust feature, 
which does not depend upon tuning the model to criticality. We argue that the mechanism 
determining the power law is very general, and that power-law distributions of small fluxes 
are a robust feature of the model.

Section \ref{sec: 2} defines our model, in both one and two dimensions. 
Our model includes both a \lq skeleton' of wetted bonds, and the flux of material, $\phi$, 
carried in each wetted bond. The skeleton is defined by three probabilities: a wetted 
channel continues with probability $p_1$, splits into $K$ channels with probability $p_2$, and 
terminates with probability $p_0$. The flux carried by a bond which splits 
is distributed so that a fraction $r_k$ flows into each branch, with $r_1+\ldots+r_K=1$, 
where $r_k\in[0,1]$. We emphasise cases where there are only $K=2$ branches, 
and where the two branches carry fixed fractions of the total flux, denoted  by 
$r$ and $1-r$. Accounting for the relation $p_0+p_1+p_2=1$, the model then has 
three independent parameters, $p_0$, $p_2$ and $r$.
The flux-conserving case is the $p_0=0$ subspace. The model which was 
considered in \cite{Kaw+17} is the one-dimensional, flux-conserving case.

Section \ref{sec: 3} considers the distribution of fluxes in the mass-conserving 
case in some detail. It is shown that, for small values of the flux, the probability
distribution function (PDF) of the flux is asymptotic to a 
power law described by equation (\ref{eq: 1.1}). 
We obtain an exact equation for the exponent $\alpha$ of this power law, 
in terms of some probabilities, $P_j$ which characterise the occupation of lattice sites.

The bond skeleton is characterised by the probability $f$ that a given 
bond is occupied. In section \ref{sec: 4} this is calculated under the 
assumption that the occupation of sites is statistically independent, and 
we find very good agreement between theory and numerical experiment. 
Section \ref{sec: 5} estimates the probabilities $P_j$ which define the 
equation for $\alpha$ using the same approach, and 
we compare empirically determined values for the exponent $\alpha$ 
with those obtained from both numerical and theoretical 
estimates of the $P_j$.

In section \ref{sec: 6} we consider the extent to which the independent-occupancy  
approximation is exact. We demonstrate two results which indicate that 
it is exact in the one-dimensional version of the model.  We find that, in the 
two-dimensional case, this is a very good approximation, but not exact. 

Section \ref{sec: 7} discusses the relationship between our system 
and the standard model for directed percolation. By varying the parameter 
$p_0$ we can induce a percolation transition in our model, 
which can be regarded as a consequence of large voids appearing 
between the active bonds. In sub-section  \ref{sec: 7.1} we investigate 
where the transition lies in the parameter space of our model.
Our model can be thought of in terms of a combination of 
\lq bond' and \lq site' deletion processes, and section \ref{sec: 7.1}
also discuss the relationship between our system and a model 
for mixed site and bond percolation, discussed in \cite{Tre+95}.
Sub-section \ref{sec: 7.2} presents some numerical evidence that the 
the critical exponents describing the structure of the skeleton are the same 
as for the usual directed percolation model. In section \ref{sec: 7.3}
we present some results on the distribution of void sizes.

Finally section \ref{sec: 8} summarises and discusses the implications of our 
results, including a model for the slow elution of material by percolation.

\section{Definition of the model}
\label{sec: 2}

We define discrete dynamical processes, with an iteration 
number, $j$. Incrementing $j$ can be thought of as advancing time, 
but in the physical contexts described in the Introduction, increasing 
$j$ represents moving downstream in the forced flow. Because the 
iteration index can be interpreted as a discrete time, our one or two 
dimensional systems are comparable to standard models for directed 
percolation in $1+1$ dimensions or $2+1$ dimensions, respectively.  

\subsection{One-dimensional model}
\label{sec: 2.1}

We consider a bi-partite lattice. At even iteration number $j$, only even sites are 
occupied. The dynamics always moves an occupied site by one unit
left or right, so that for odd iteration number, only the odd sites may be occupied.
Every occupied site, index $i$, is associated with a flux, $\phi_i(j)$, at iteration $j$.

First consider the dynamics of the site occupations. At each iteration, an 
occupied site is annihilated with probability $p_0$, or moves either 
right or left with equal probabilities $p_1/2$, or else it splits into $K=2$ branches 
with probability $p_2$.

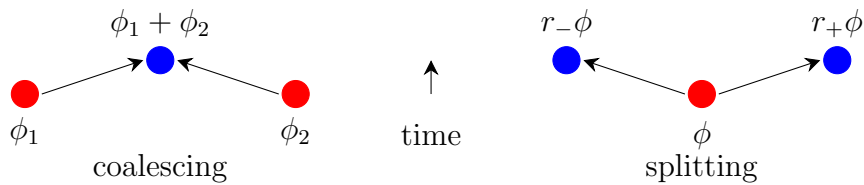
\begin{figure}
\centering
\begin{tikzpicture}[scale=0.9]
\draw[fill,red] (1,1.1) circle (0.2);
\draw[fill,blue] (3,1.6) circle (0.2);
\draw[fill,red] (5,1.1) circle (0.2);
\draw[fill,blue] (9,1.6) circle (0.2);
\draw[fill,red] (11,1.1) circle (0.2);
\draw[fill,blue] (13,1.6) circle (0.2);
\node[] at (1,0.55) {$\phi_1$};
\node[] at (5,0.55) {$\phi_2$};
\node[] at (3,2.15) {$\phi_1+\phi_2$};
\node[] at (11,0.5) {$\phi$};
\node[] at (9,2.15) {$r_-\phi$};
\node[] at (13,2.15) {$r_+\phi$};
\node[] at (3,0) {${\rm coalescing}$};
\node[] at (11,0) {${\rm splitting}$};
\draw[-{Stealth[length=2mm, width = 2mm]}] (1.25,1.1) -- (2.75,1.6) node[pos=.5,left] {};
\draw[-{Stealth[length=2mm, width = 2mm]}] (4.75,1.1) -- (3.25,1.6) node[pos=.5,left] {};
\draw[-{Stealth[length=2mm, width = 2mm]}] (10.75,1.1) -- (9.25,1.6) node[pos=.5,left] {};
\draw[-{Stealth[length=2mm, width = 2mm]}] (11.25,1.1) -- (12.75,1.6) node[pos=.5,left] {};
\draw[-{Stealth[length=2mm, width = 2mm]}] (7.0,1.1) -- (7.0,1.6) node[pos=.5,left] {};
\node[] at (7,0.5) {${\rm time}$};
\end{tikzpicture}
\caption{
The model is defined on a bi-partite lattice. Only even sites (red) are 
occupied at even-numbered iterations, only odd sites (blue) during odd iterations.
At each iteration every occupied site is either removed (probability $p_0$), 
moves to one of its nearest neighbours (probability $p_1$), or else splits into $K=2$ 
daughters (probability $p_2$). The flux $\phi$ carried by an occupied site moves with it, 
unless the site is removed, in which case the flux disappears. When two fluxes are moved to 
the same site, they are added.
}
\label{fig: 2.1}
\end{figure}

The corresponding dynamics of the fluxes is as follows. If a site is annihilated 
its flux disappears.  When transitions bring two occupied sites to the same position, 
their fluxes combine, 
and if the site branches, then its flux is divided. So values of the flux are changed 
by two possible processes:
\begin{eqnarray}
\label{eq: 2.1}
{\rm coalescence}:&\quad\quad&\phi_i(j+1)=\phi_{i-1}(j)+\phi_{i+1}(j)
\nonumber \\
\ \ \ \ \ {\rm splitting}:&\quad\quad&\phi_{i-1}(j+1)=r_-\phi_{i}(j)
\ ,\ \ \ \phi_{i+1}(j+1)=r_+\phi_{i}(j)
\end{eqnarray}
where either $(r_+,r_-)=(r,1-r)$ or $(r_+,r_-)=(1-r,r)$, both cases with probability 
equal to one-half. The model is illustrated schematically in figure \ref{fig: 2.1}.
More generally, we can make $r\in[0,1]$ a random variable, chosen independently for 
every bifurcation, with a PDF which is symmetric about $\frac{1}{2}$. 

The model only describes a volume-preserving flow when the stopping probability $p_0$ is equal
to zero. This parameter is included for two reasons. 
The parameter $p_0$ is included 
primarily so that our system encompasses the standard model of directed 
percolation. In addition, the case where $p_0>0$ can model 
situations where the solvent disappears, for example by evaporation.

The fluxes at each wetted site can increase (due to coalescence) or decrease 
(due to splitting), and some sites become unoccupied. The long-time behaviour of 
the model is characterised by the probability, $f$, that sites are occupied and the distribution 
of the non-zero values of $\phi$.  A nonzero value of $p_0$ implies a loss of 
conservation, which significantly changes the nature of the problem.
When $p_0$ is large enough, the probability that a site
is occupied at long times becomes zero, and as $p_0$ decreases, one observes a transition when
the probability of occupation becomes strictly positive. 
 
 \subsection{Two-dimensional model}
\label{sec: 2.2}

Consider a model defined on an $2N\times 2N$ square lattice, with sides 
identified to make a toroidal topology. Every site $(i,j)$ carries a weight $\phi_{ij}$. 
Initially only sites with $i+j$ even are occupied, with unit weight, so that our model 
is defined on a bi-partite lattice.  The value of $N$ is assumed to be large.

The lattice configuration is then evolved in discrete timesteps. At each site, 
we choose (with equal probabilities) a move to one of the four nearest 
neighbour sites. With probability $p_1$, all of the flux on $(i,j)$ is moved to 
the selected neighbour. Alternatively, with probability $p_2$, branching 
onto $K$ nearest neighbours occurs. Note that these moves preserve
the bi-partite property, so that at iteration $k$, only sites with $i+j$ having the 
same parity as $k$ may be occupied.

We investigated two different versions of this model, which we term 
the \emph{two-branch} and \emph{four-branch} models.  
In the two-branch model, we have $K=2$, and a fraction $1-r$ of the weight 
at $(i,j)$ is moved to the selected neighbour, and a fraction $r$ is moved in the 
reciprocal direction. All of the random choices are independent. 
The model has three parameters of interest, $p_2$, $p_0$ and $r$. 

In the four-branch model, there is branching to all of the nearest 
neighbour sites, so that $K=4$. The fractional weights for each branch 
are defined by four numbers, $\{r_1,r_2,r_3,r_4\}$, with $r_1+r_2+r_3+r_4=1$. 

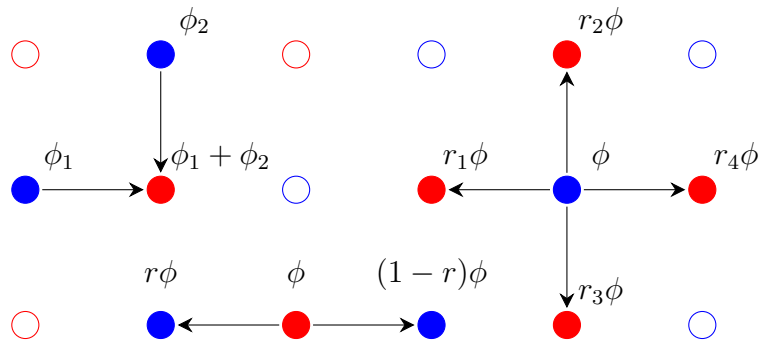
\begin{figure}
\centering
\begin{tikzpicture}[scale=0.9]
\draw[red] (1,0) circle (0.2);
\draw[fill,blue] (1,2) circle (0.2);
\draw[red] (1,4) circle (0.2);
\draw[fill,blue] (3,0) circle (0.2);
\draw[fill,red] (3,2) circle (0.2);
\draw[fill,blue] (3,4) circle (0.2);
\draw[fill,red] (5,0) circle (0.2);
\draw[blue] (5,2) circle (0.2);
\draw[red] (5,4) circle (0.2);
\draw[fill,blue] (7,0) circle (0.2);
\draw[fill,red] (7,2) circle (0.2);
\draw[blue] (7,4) circle (0.2); 
\draw[fill,red] (9,0) circle (0.2);
\draw[fill,blue] (9,2) circle (0.2);
\draw[fill,red] (9,4) circle (0.2);
\draw[blue] (11,0) circle (0.2);
\draw[fill,red] (11,2) circle (0.2);
\draw[blue] (11,4) circle (0.2);   
\node[] at (1.5,2.5) {$\phi_1$};
\node[] at (3.5,4.5) {$\phi_2$};
\node[] at (3.5,2.5) {$\ \ \ \ \ \phi_1+\phi_2$};
\node[] at (5,0.75) {$\phi$};
\node[] at (3,0.75) {$r\phi$};
\node[] at (7,0.75) {$(1-r)\phi$};
\node[] at (9.5,2.5) {$\phi$};
\node[] at (9.5,4.5) {$r_2\phi$};
\node[] at (9.5,0.5) {$r_3\phi$};
\node[] at (7.5,2.5) {$r_1\phi$};
\node[] at (11.5,2.5) {$r_4\phi$};
\draw[-{Stealth[length=2mm, width = 2mm]}] (1.25,2) -- (2.75,2) node[pos=.5,left] {};
\draw[-{Stealth[length=2mm, width = 2mm]}] (3,3.75) -- (3,2.25) node[pos=.5,left] {};
\draw[-{Stealth[length=2mm, width = 2mm]}] (5.25,0) -- (6.75,0) node[pos=.5,left] {};
\draw[-{Stealth[length=2mm, width = 2mm]}] (4.75,0) -- (3.25,0) node[pos=.5,left] {};
\draw[-{Stealth[length=2mm, width = 2mm]}] (9.25,2) -- (10.75,2) node[pos=.5,left] {};
\draw[-{Stealth[length=2mm, width = 2mm]}] (8.75,2) -- (7.25,2) node[pos=.5,left] {};
\draw[-{Stealth[length=2mm, width = 2mm]}] (9,1.75) -- (9,0.25) node[pos=.5,left] {};
\draw[-{Stealth[length=2mm, width = 2mm]}] (9,2.25) -- (9,3.75) node[pos=.5,left] {};
\end{tikzpicture}
\caption{
In two dimensions, when the flux at an occupied site splits 
there may be up to four branches. We consider two cases in some detail.
In the  \emph{two-branch} version of the model there are two branches, 
which go in opposite directions, either horizontal or vertical, with equal probability.
In the \emph{four-branch} model, where branching events reach all nearest neighbours, 
with the locations of the weights $r_1,r_2,r_3,r_4$ randomly assigned.
}
\label{fig: 1}
\end{figure}

\section{Power-law distribution of fluxes in the mass-conserving case}
\label{sec: 3}

In the following, Sections~\ref{sec: 3}, \ref{sec: 4}, 
\ref{sec: 5} and \ref{sec: 6}, we consider exclusively the mass conserving 
case, with $p_0 = 0$. To simplify the notation, we will set, in these sections, $ p = p_2$, 
so that $p_1 = (1 - p)$. 

We argue here that the  
distribution of fluxes $P(\phi)$ has a power-law behaviour in the limit as 
$\phi\to 0$ (with the exponent in (\ref{eq: 1.1}) satisfying $\alpha<1$, so that the distribution is normalisable).
Branching of a channel reduces the flux due to multiplying by a 
random factor $r_k<1$, which we assume to  
have a known PDF. Because fluxes are added when coalescence of channels 
occurs, this process increases the flux. We are interested in the distribution 
of very small values of the flux $\phi$. In this case splitting and coalescence have 
very different effects. In the case where 
a channel carries a very small flux, coalescence with another channel will 
produce a much larger flux (with a value which is typically comparable to the 
mean flux). Almost all coalescence events will, therefore, remove a very small value of $\phi$, 
whereas splitting events just reduce its value. The distribution of very small 
values of $\phi$ is, therefore, the result of a competition between two process: the 
small values of $\phi$ continue to decrease due to splitting, but they are annihilated 
by coalescences. 

In order to explain why a power-law distribution of the flux is expected, 
we start by making a change of variables. Instead of considering
$\phi$, we consider the probability density function (PDF) of a 
logarithmic variable, $\psi\equiv \ln\,\phi$. 
Consider the dynamics of the variable $\psi$ (regarding increasing $\psi$ as
a displacement to the right). With every bifurcation of a channel, the points
representing the values of $\psi$ are split into $K$ new points, and 
each one is displaced by $\ln\,r_k$. 
When two channels coalesce, the two values of
$\psi$ are replaced by $\psi=\ln\,[\exp(\psi_1)+\exp(\psi_2)]$. 
In the following we shall assume that the PDF $P(\phi)$ is bounded so that
the probability of $\phi$ being less than $\phi_0$ approaches zero as $\phi_0\to 0$.
This is consistent with the distribution (\ref{eq: 1.1}) provided $\alpha<1$. Under this assumption, in
the limit as $\phi \to 0$, most coalescences occur with channels carrying a much larger flux. 
As a consequence,  coalescence of a channel with a small flux, $\psi$, is replaced 
by a value close to that which characterises a typical channel. This picture implies that 
the variable $\psi$ drifts to the left with each bifurcation, but, in the case 
of small fluxes, coalescence almost inevitably causes a jump back to a 
position close to the origin. 

\begin{figure}
	\includegraphics[width=8.2cm]{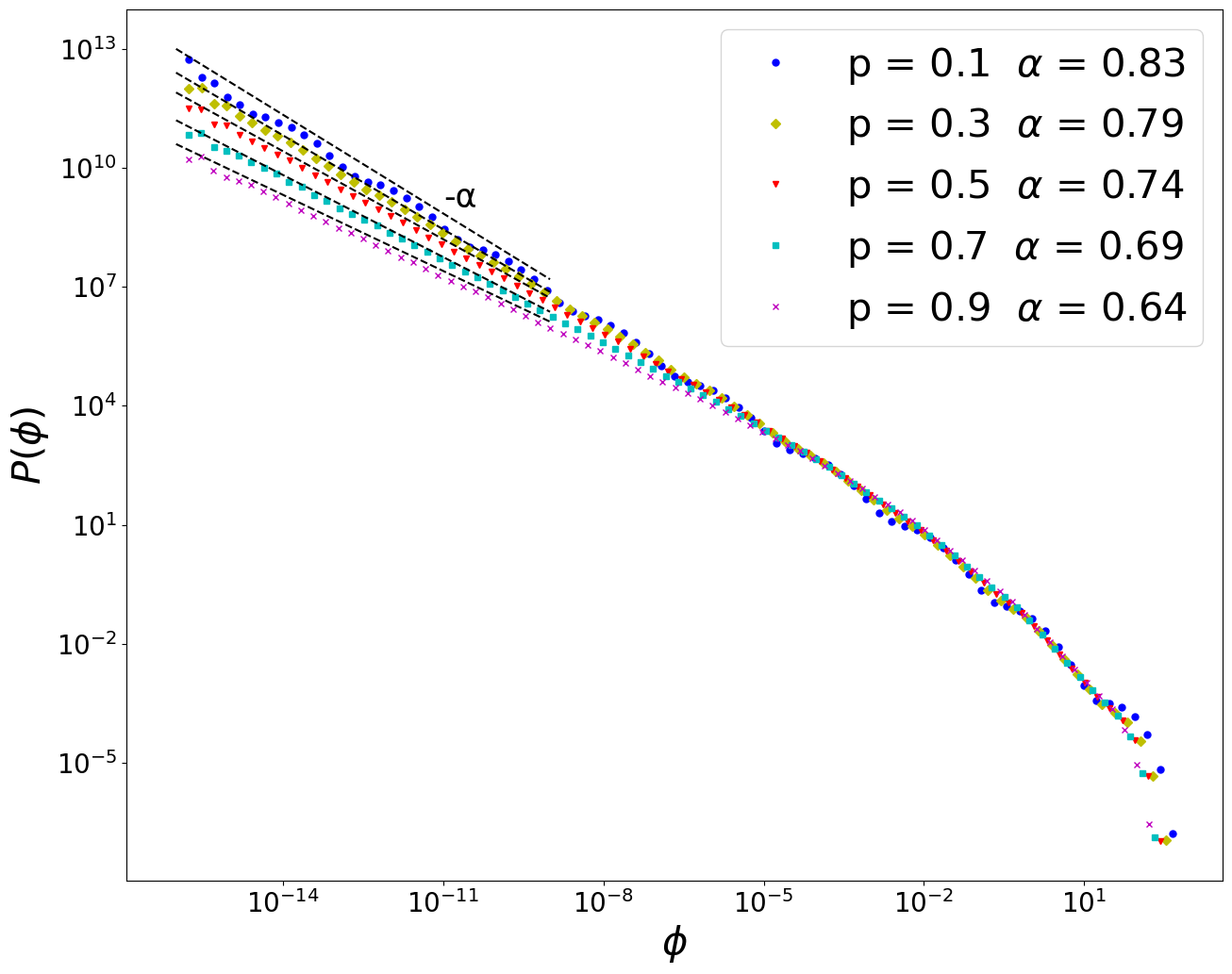}
	\includegraphics[width=8.2cm]{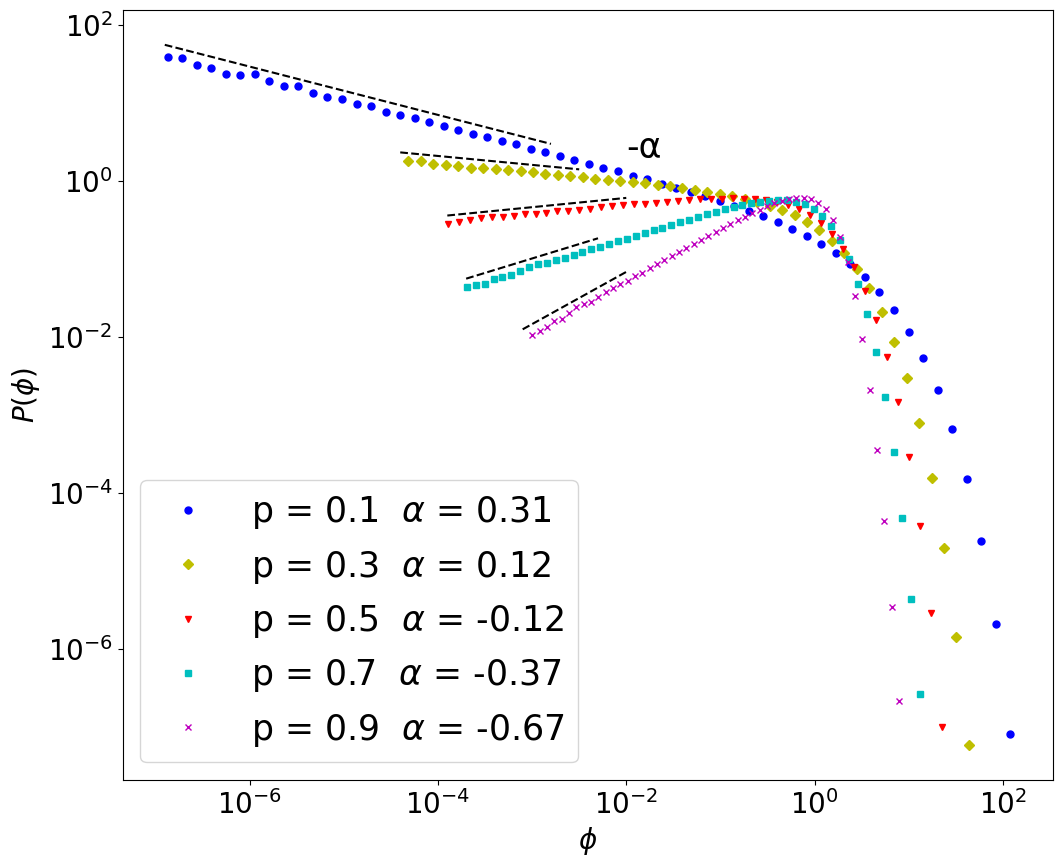}
	\includegraphics[width=8.2cm]{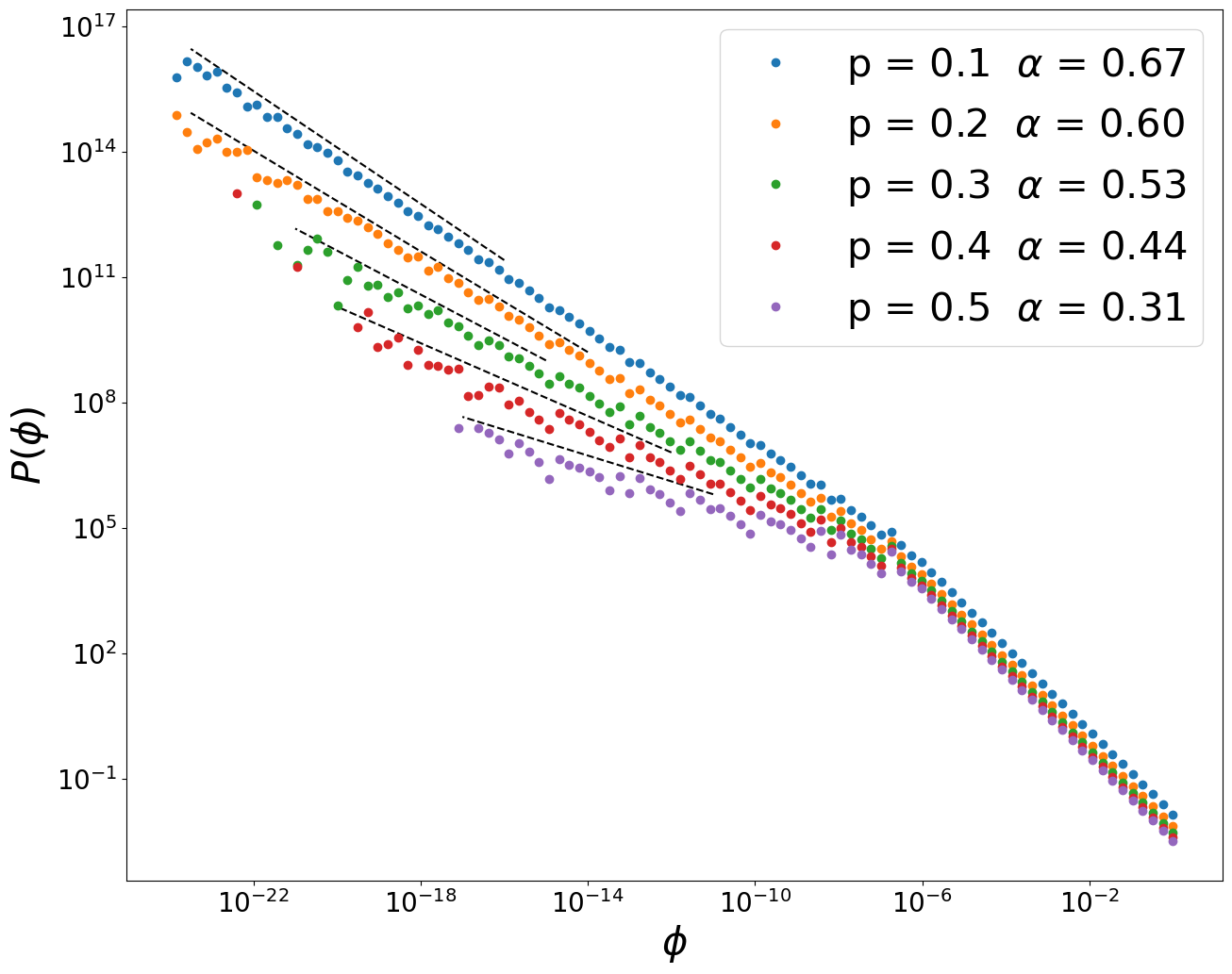}
	\includegraphics[width=8.2cm]{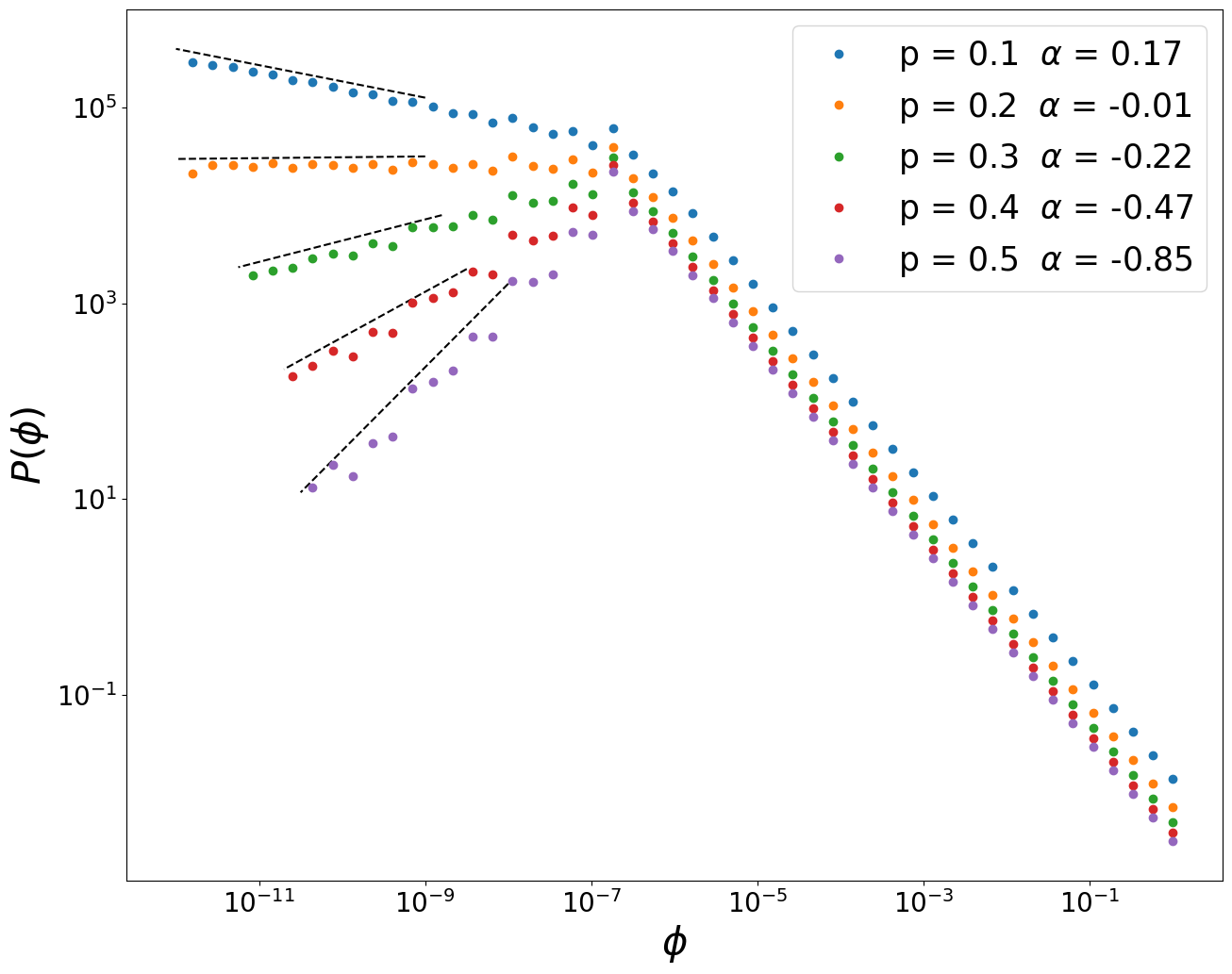}	
\caption{
\label{fig: 3.1}
Results of numerical simulations demonstrating that the distribution of flux, $P(\phi)$, is 
asymptotic to a power law at small $\phi$. We show results for the two-dimensional model. 
Upper row: two-branch system: each panel illustrates different values of $p$, with 
$r=0.01$ (left panel) and $r=0.5$ (right panel). 
Lower row: four-branch system, with $r=0.025$ (left panel) and $r=0.25$ (right panel).
}
\end{figure}

Because the equations defining the dynamics of $\psi$ become independent 
of the value of $\psi$ in the limit as $\psi \to -\infty$, the PDF of $\psi$ should reflect this translational
symmetry. In the limit as $\psi\to -\infty$, 
the PDF of $\psi$ should be asymptotic to an eigenfunction of the translation 
operator. Because the exponential function is an eigenfunction of 
a translation operator, we expect that the PDF of $\psi$ has the form 
\begin{equation}
\label{eq: 3.1}
P_\psi(\psi)\sim \exp(\lambda \psi)
\ .
\end{equation}
Note that we must have $\lambda>0$ to have a normalisable distribution
if this law holds as $\psi\to -\infty$. The corresponding distribution of
$\phi$ is then a power law of the form 
\begin{equation}
\label{eq: 3.2}
P(\phi)\sim \phi^{-\alpha},\ \ \ \  \alpha=1-\lambda 
\ . 
\end{equation}
This is a very general argument indicating that the steady-state 
distribution of fluxes approaches a power law as we go deeper into the 
percolation medium, but it does not yield a prediction of the 
exponent $\alpha$. Figure \ref{fig: 3.1} illustrates numerical 
simulations of the distribution $P(\phi)$ for the two-dimensional model, 
demonstrating that it is indeed asymptotic to a power law at small values of $\phi$.

To calculate $\alpha$ we shall determine a master equation for the PDF 
of the variable $\psi$, valid in the limit as $\psi\to -\infty$. Because 
coalescence almost
inevitably results in the value of $\psi$ making a large jump to the right, we 
must consider the fate of sites which are occupied \emph{and which 
have not experienced a coalescence for a large number of iterations}. The latter requirement
is imposed because it is only those sites which have very small values of $\phi$. For this
subset of sites we introduce the following probabilities:
\begin{eqnarray}
\label{eq: 3.3}
P_1&\equiv&{\rm \ probability\ of\ moving\ without\ branching\ or\ coalescence}
\nonumber \\
P_2&\equiv&{\rm \ probability\ of\ moving\ without\ branching\ and\ undergoing\ coalescence}
\nonumber \\
P_{3+k}&\equiv&{\rm \ probability\ of\ branching\ }K{\rm\ ways,\ with\ }k
{\rm\ branches\ undergoing\ coalescence}
\nonumber \\
\end{eqnarray}
Note that $P_1+P_2=1-p$ and $P_3+\ldots+P_{3+K}=p$. 
In practice, when we estimate the $P_k$ from numerical simulations, 
we only accumulate statistics for those sites which have not undergone 
coalescence in the preceding $N_{\rm thr}$ iterations. We take a sufficiently 
large value for this threshold such that the $P_k$ are insensitive to the value $N_{\rm thr}$.

To describe the asymptotic form of the probability of very small fluxes, 
we write down an equation for the PDF of $\psi$ at iteration $j+1$, which is 
valid in the limit as $\psi\to -\infty$. Note that, because events involving coalescence 
almost always induce a large increase of the flux, they do not contribute to this 
balance equation for very small values of $\phi$. It follows that it is only events 
which do not involve coalescence at iteration $j$ which contribute to 
$P_\psi(\psi,j+1)$, when $\psi\to -\infty$: these events are moving without coalescence, 
leaving $\psi$ unchanged (probability $P_1$), or splitting events where some of the daughters
escape coalescence (probabilities $P_3,\ldots,P_{2+K}$). Taking this remark into account, if 
$P_\psi(\psi,j)$ is the PDF of $\psi$ at iteration $j$, then
\begin{equation}
\label{eq: 3.4}
P_\psi(\psi,j+1)=P_1P_\psi(\psi,j)
+Q\sum_{k=1}^K P_\psi(\psi -\ln(r_k),j) 
\end{equation}
with 
\begin{equation}
\label{eq: 3.5}
Q=\frac{1}{K}\left[KP_3+(K-1)P_4+\ldots+P_{K+2}\right]
\ .
\end{equation}
Seeking a solution of the form (\ref{eq: 3.1}) which is independent of $j$ gives 
an \emph{exact} equation for the exponent $\alpha$:
\begin{equation}
\label{eq: 3.6}
1-P_1=Q\sum_{k=1}^K r_k^{\alpha-1}
\ .
\end{equation}
In the cases where $K=2$ and where the splitting ratios are $(r,1-r)$), (this includes the 
one-dimensional model and the two-branch model in two dimensions) equation 
(\ref{eq: 3.6}) simplifies to 
\begin{equation}
\label{eq: 3.7}
r^{\alpha-1}+(1-r)^{\alpha-1}= F(p)
\end{equation}
with
\begin{equation}
\label{eq: 3.8}
F(p) \equiv  \frac{1-P_1}{Q}=\frac{2(1-P_1)}{2P_3+P_4}
\ \ \ \ \ \ ({\rm {\bf two-way\ splitting}})
\ .
\end{equation}
We also simulated a model with $K=4$: when there is a four-way split, 
we set a weight factor of $r$ for two of the sites (chosen at random), and a factor
$1-r$ for the other two sites. In this case, $\alpha$ satisfies equation (\ref{eq: 3.7}) with
\begin{equation}
\label{eq: 3.9}
F(p) \equiv  \frac{1-P_1}{2Q}=\frac{(1-P_1)}{2P_3+3P_4/2+P_5+P_6/2}
\ \ \ \ \ \ ({\rm {\bf four-way\ splitting}})
\ .
\end{equation}

\section{Occupation probabilities in the mass-conserving case}
\label{sec: 4}

Here we present calculations for the occupation probability $f$ in the 
mass-conserving case, $p_0 = 0$, assuming that the 
occupation of sites is statistically independent of their neighbours. 
For the sake of simplicity, we set $p = p_2$, so $p_1 = 1 - p$.
We limit the discussion to the two-dimensional models, because the 
one-dimensional case was treated in \cite{Kaw+17}, 
where it was shown that the occupation probability is 
\begin{equation}
\label{eq: 4.1}
\fo(p) = \frac{4p}{(1+p)^2}
\ ,
\end{equation}
where the sub-index 1 indicates the one-dimensional model.

\subsection{Occupation probability: two-dimensional case with double branching}
\label{sec: 4.1}

\begin{figure}
	\includegraphics[width=10cm]{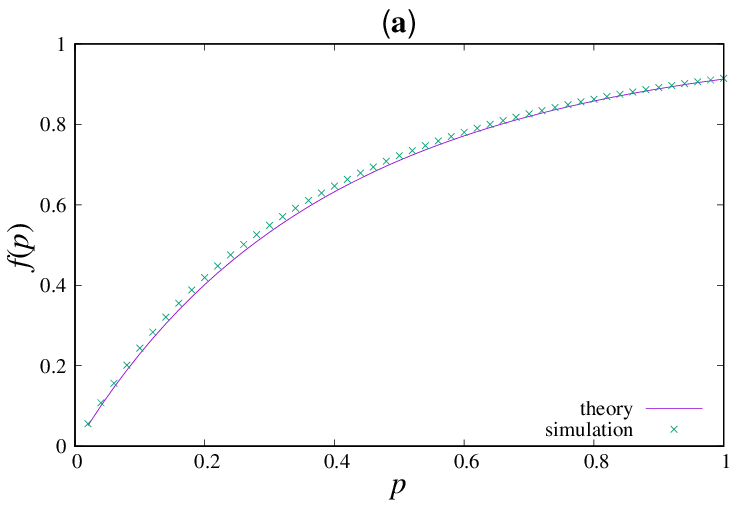}
	\includegraphics[width=6cm]{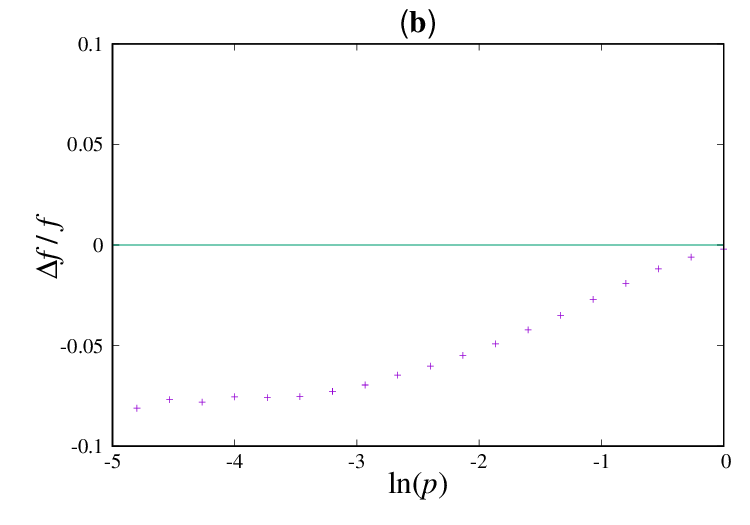}
\caption{
\label{fig: 4.1}
({\bf a}) Comparing simulated occupation fraction $f_{2,2}$ for the 
two-branch model with prediction from 
independent-occupancy approximation, equations (\ref{eq: 4.1.5}) and (\ref{eq: 4.1.6}). 
({\bf b}) Shows the fractional error, plotted against $\ln p$. 
(We defined the fractional error by 
$\Delta f/f\equiv (f_{\rm th}-f_{\rm num})/\sqrt{f_{\rm th}f_{\rm num}}$, where
$f_{\rm th}$ and $f_{\rm num}$ are, respectively, the theoretical and numerically determined 
values of $f$.) 
}
\end{figure}

We determine here the probability of occupation
in the two-dimensional problem, first
with $K = 2$ (\emph{two-branch} model), $\ftt(p)$, where the sub-indices indicate 
the two-dimensional model with two branches and we recall that $p = p_2$. 
Assuming that sites are randomly occupied with probability $\ftt$,
we estimate the probability $P_{\rm empty}$ that a site will be 
empty at the next iteration. Note that the probability that one of the four 
nearby sites makes a transition to reach this site is 
\begin{equation}
\label{eq: 4.1.1}
P_{\rm tr}=\frac{(1+p)}{4}\ftt
\ .
\end{equation}
The probability of the site remaining empty is 
\begin{equation}
\label{eq: 4.1.2}
P_{\rm empty}=\left(1-P_{\rm tr}\right)^4=1-\ftt
\ .
\end{equation}
This leads to a cubic equation for $\ftt$:
\begin{eqnarray}
\label{eq: 4.1.3}
1-\ftt&=&\left[1-\left(\frac{1+p}{4}\right)\ftt \right]^4 
\nonumber \\
&=&1-(1+p)\ftt+\frac{6}{16}(1+p)^2\ftt^2
\nonumber \\
&&-\frac{1}{16}(1+p)^3\ftt^3+\frac{1}{256}(1+p)^4\ftt^4\ .
\end{eqnarray}
When $p\ll 1$, this is approximated by 
$-p\ftt+\frac{3}{8}\ftt^2\sim 0$, so that 
\begin{equation}
\label{eq: 4.1.4}
\ftt \sim \frac {8}{3}p
\ .
\end{equation}
The cubic equation arising from (\ref{eq: 4.1.3}) can, in fact, be solved by the method of Cardano.
Within the interval $p\in(0,1)$, we have only one real root.
By this method, the dependence of $\ftt$ on $p$ is given by:
\begin{equation}
\label{eq: 4.1.5}
\ftt (p)=\frac{4}{3(1+p)}\left[4-\left(\frac{y(p)-x(p)}{1+p}\right)^{1/3}+
\left(\frac{x(p)+y(p)}{1+p}\right)^{1/3}\right]\ ,
\end{equation}
where
\begin{equation}
\label{eq: 4.1.6}
x(p)=44-10p
\ ,\ \ \ 
y(p)=6\sqrt{3}\sqrt{p^{2}-8p+18}
\ .
\end{equation}
The maximum of $\ftt(p)$ is $\ftt \approx 0.9126$ at $p = 1$.
Figure \ref{fig: 4.1}(a) compares this prediction 
of the filling probability with the result of numerical simulation. The 
agreement is very good, but not perfect. In particular, we
find that the fractional error is quite large at small values of $p$, 
as illustrated in panel \ref{fig: 4.1}(b). We were not able to determine
whether the fractional error eventually approaches zero as $p\to 0$.

\subsection{Occupation probability: two-dimensional case with fourfold branching}
\label{sec: 4.2}

\begin{figure}
        \includegraphics[width=10cm]{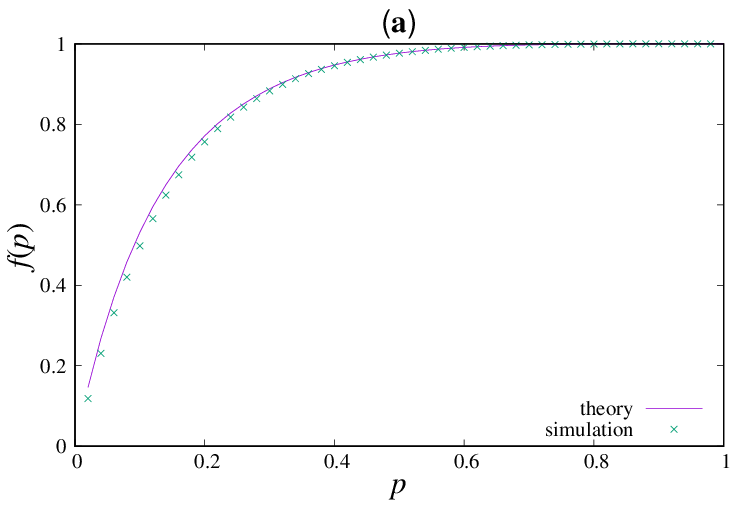}
	\includegraphics[width=6cm]{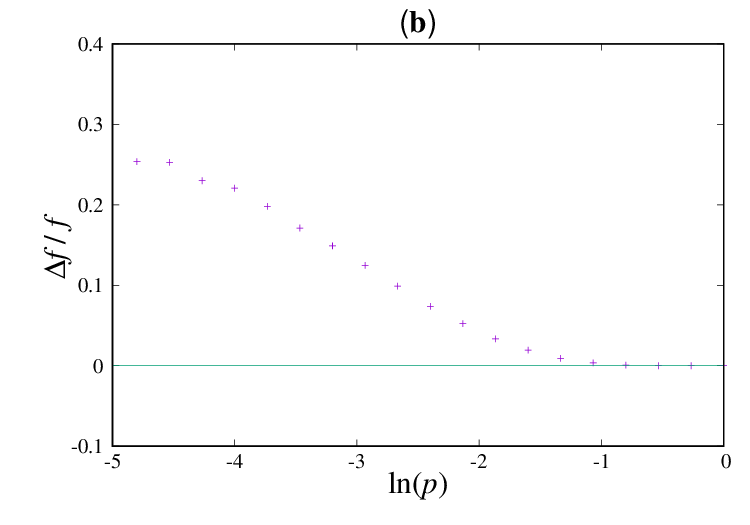}
\caption{
\label{fig: 4.2}
({\bf a}) Comparing simulated occupation probability $f_{2,4}$ for the 
four-branch model with prediction from 
independent-occupancy approximation, equation (\ref{eq: 4.2.3}). 
({\bf b}) Shows the fractional error (defined in the same way as for figure \ref{fig: 4.1}), 
plotted against $\ln p$.
}
\end{figure}

The occupation probability for the four-branch model in two dimensions
will be denoted $f_{2,4}$. For this model the transition probability is
\begin{equation}
\label{eq: 4.2.1}
P_{tr}=\frac{1+3p}{4}f_{2,4}
\ .
\end{equation}
Then, the cubic equation for $f_{2,4}$ is:
\begin{equation}
\label{eq: 4.2.2}
1-f_{2,4}=\left[1-\left(\frac{1+3p}{4}\right)f_{2,4}\right]^{4}
\ .
\end{equation}
Making a comparison with equation (\ref{eq: 4.1.3}), we see that 
$f_{2,4}(p)=f_{2,2}(3p)$, so that the occupation probability in this case is
\begin{equation}
\label{eq: 4.2.3}
f_{2,4}(p)=\frac{4}{3(1+3p)}
\left[4-\left(\frac{y(3p)-x(3p)}{1+3p}\right)^{1/3}+\left(\frac{x(3p)+y(3p)}{1+3p}\right)^{1/3}\right]\ ,
\end{equation}
and, when $p\ll 1$, we have $f_{2,4}\sim 8p$. In this case, the maximum value is $f_{2,4}=1$ at $p=1$.
Figure \ref{fig: 4.2} compares this prediction of the filling probability with the 
result of numerical simulation. Again, while the agreement is very good throughout most of the 
range of $p$, there is a substantial fraction error for small values of $p$.

\section{Estimates of transition probabilities in the mass-conserving case}
\label{sec: 5}

In section \ref{sec: 3} we presented an exact equation, (\ref{eq: 3.6}), determining 
the exponent $\alpha$ in terms of a set of probabilities $P_k$, defined by equation 
(\ref{eq: 3.3}). In this section we consider these probabilities, comparing 
numerical simulations with theoretical estimates, where these are available.

We consider different models in turn (including the one-dimensional model,
because these probabilities were not given in \cite{Kaw+17}). In all cases, the 
theory uses the assumption that the sites are independently 
occupied with probability $f$, as estimated in section \ref{sec: 4}. 
In section \ref{sec: 6} we shall argue that this independent-occupation 
assumption is exact in the one-dimensional case. 
Accordingly we propose that the formulae for the $P_k$ are exact in one dimension.
 
In the case of the two-dimensional model with two-way splitting, we are able to 
estimate the $P_k$ analytically using the independent-occupation model. We find
close agreement with values derived from numerical simulations. In the two-dimensional
model with four-way splitting, we are limited to giving values of $P_k$ derived from 
simulations.

\subsection{One-dimensional model}
\label{sec: 5.1}

A given trail can evolve in several ways at each iteration, with probabilities defined
by equation (\ref{eq: 3.3}).
If sites are occupied with probability $\fo (p)$, 
we find, using Eq.~(\ref{eq: 4.1}), that there
is a \emph{transition probability} for 
a trail coalescing with
one or other of its two neighbouring sites, given by
\begin{equation}
\label{eq: 5.1.1}
P_{\rm tr}=\left[\frac{1}{2}(1-p)+p\right]\fo(p)=\frac{2p}{1+p}
\end{equation}
(the term $(1-p)/2$ comes from the case where the neighbouring site does not divide and 
moves in the direction that creates a collision, and $p$ comes from the case where
the neighbouring trail divides).

The probability $P_1$ arises from the case where a trail does not divide, and does not collide:
\begin{equation}
\label{eq: 5.1.2}
P_1=(1-p)(1-P_{\rm tr})=\frac{(1-p)^2}{1+p}
\ .
\end{equation}
Similarly
\begin{equation}
\label{eq: 5.1.3}
P_2=(1-p)P_{\rm tr}=\frac{2p(1-p)}{1+p}
\ .
\end{equation}
In the case where the trail divides, there are two independent chances for the trail 
to be annihilated, so the probability for both daughter trails to end is
\begin{equation}
\label{eq: 5.1.4}
P_5=p[P_{\rm tr}]^2=\frac{4p^3}{(1+p)^2}
\ .
\end{equation}
Similarly the probability for both daughter trails to survive is 
\begin{equation}
\label{eq: 5.1.5}
P_3=p[1-P_{\rm tr}]^2=\frac{p(1-p)^2}{(1+p)^2} 
\ .
\end{equation}
And because there are two ways in which one daughter trail can continue
\begin{equation}
\label{eq: 5.1.6}
P_4=pP_{\rm tr}[1-P_{\rm tr}]=\frac{4p^2(1-p)}{(1+p)^2}
\ .
\end{equation}
Using equations (\ref{eq: 5.1.2}) to (\ref{eq: 5.1.6}) in equation (\ref{eq: 3.7}), we 
find that for the one-dimensional flux-conserving model, $\alpha$ is a solution of
\begin{equation}
\label{eq: 5.1.7}
r^{\alpha-1}+(1-r)^{\alpha-1}=\frac{3-p}{1-p}
\end{equation}
in accord with Equation (2) of \cite{Kaw+17}.

\subsection{Two-dimensional model with two branches}
\label{sec: 5.2}

Next consider estimates for the transition probabilities for the two-dimensional 
model with branching into two opposite directions. Again, we use the assumption that 
the sites are independently occupied, with probability $\ftt$, as approximated 
by equations (\ref{eq: 4.1.5}), (\ref{eq: 4.1.6}).

\begin{figure}
	\includegraphics[width=10cm]{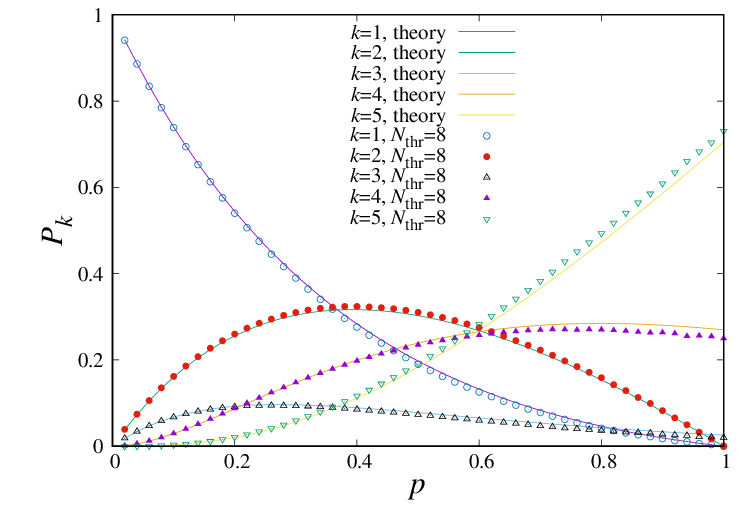}
\caption{
\label{fig: 5.1}
Estimates of the probabilities defined in (\ref{eq: 3.3}), $P_k(p)$, $k=1,\ldots,5$, for the 
two-dimensional two-branch model. The solid lines are the theoretical predictions, 
equations (\ref{eq: 5.2.1}) to (\ref{eq: 5.2.5}). The numerical simulations imposed the 
requirement that the site has not recently experienced a coalescence in the preceding 
$N_{\rm thr}=8$ iterations. They differ significantly from the theoretical model as $p\to 1$.
}
\end{figure}

First note that a site moves to one of its nearest neighbours with 
a \emph{transition probability}, given by equation (\ref{eq: 4.1.1}), namely
$P_{\rm tr}=\ftt(p) ( (1-p)/4 + p/2) = \ftt(p) (1 + p)/4$.
 A site remains un-branched with probability $(1-p)$, and un-combined 
if there is no other transition into the final site from any of its three other 
neighbours. Hence 
\begin{equation}
\label{eq: 5.2.1}
P_1=(1-p)(1-P_{\rm tr})^3
\ .
\end{equation}
The value of $P_2$ is then determined by noting that $1-p=P_1+P_2$:
\begin{equation}
\label{eq: 5.2.2}
P_2=1-p-P_1
\ .
\end{equation}
If the trajectory branches (with probability $p$), it avoids collision if 
none of the three nearest neighbours of each of the two 
new sites make a transition which lands there. Hence
\begin{equation}
\label{eq: 5.2.3}
P_3=p(1-P_{\rm tr})^6
\ .
\end{equation}
Similarly, $P_5$ is the probability of an event where neither 
of the two branches avoids coalescence with at least on one of its three 
nearest neighbours:  
\begin{equation}
\label{eq: 5.2.4}
P_5=p\left[1-(1-P_{\rm tr})^3\right]^2
\ .
\end{equation}
To determine $P_4$ we can use $P_3+P_4+P_5=p$ to obtain
\begin{equation}
\label{eq: 5.2.5}
P_4=2p(1-P_{\rm tr})^3\left[1-(1-P_{\rm tr})^3\right]
\ .
\end{equation}
Figure \ref{fig: 5.1} compares these theoretical estimates for the $P_k$ with 
numerical simulations. The numerical simulations include 
only sites which had not experienced coalescence in the preceding 
$N_{\rm thr}=8$  iterations (the parameter $N_{\rm thr}$ was introduced in the 
paragraph below equation (\ref{eq: 3.3})). These simulations show that the 
true $P_k$ values differ slightly from 
the theoretical expressions, equations (\ref{eq: 5.2.1})-(\ref{eq: 5.2.5}), 
which are plotted in figure \ref{fig: 5.1}. 
The small difference between theory and simulation becomes 
negligible as $p\to 0$.  

We used the theoretical expression
for $f(p)$, equations (\ref{eq: 4.1.5}), (\ref{eq: 4.1.6}), when evaluating 
equations (\ref{eq: 5.2.1})-(\ref{eq: 5.2.5}) for figure \ref{fig: 5.1}.
Plotting the $P_k$ for simulations with $N_{\rm thr}=0$ yields curves which are barely 
distinguishable from equations (\ref{eq: 5.2.1})-(\ref{eq: 5.2.5}), indicating 
that the small discrepancy is due to the theory neglecting the requirement to 
exclude sites which have undergone recent collisions, rather than the error in equations 
(\ref{eq: 4.1.5}), (\ref{eq: 4.1.6}).

It is impractical to impose 
very large values of $N_{\rm thr}$ because, as $p\to 1$, very few sites satisfy 
the requirement to have undergone no coalescences in the last $N_{\rm thr}$ 
iterations: for $N_{\rm thr}=8$, we found that the probability of an occupied site
satisfying this criterion falls from $0.64$ at $p=0.05$ to $0.0013$ at $p=0.75$. 
Simulations with $N_{\rm thr}=4$ gave points which appear coincident with those 
for $N_{\rm thr}=8$ when included in figure \ref{fig: 5.1}.

\begin{figure}
	\includegraphics[width=8.2cm]{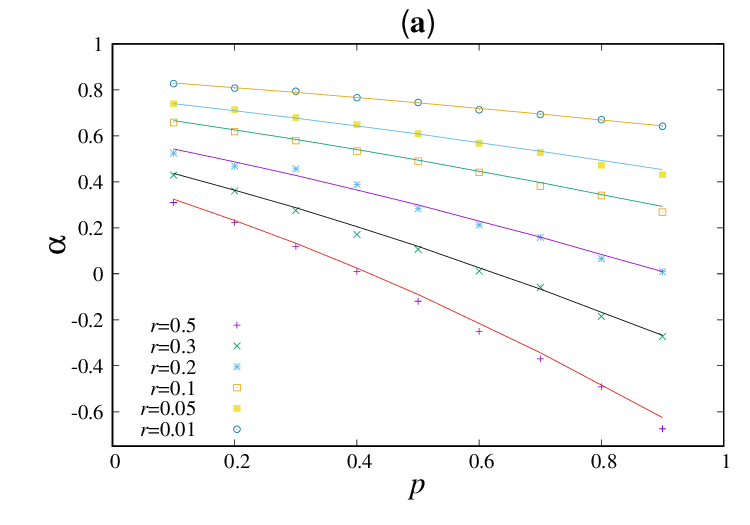}
		\includegraphics[width=8.2cm]{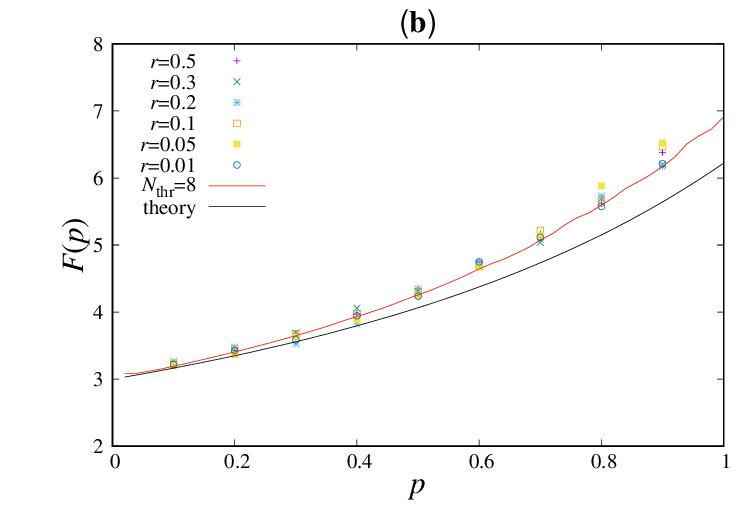}
\caption{
\label{fig: 5.2}
Testing the determination of $\alpha$ for the two-branch two-dimensional, two-branch 
model, using equation (\ref{eq: 3.6}), and the $P_k$ derived from simulations 
with $N_{\rm thr}=8$. Left panel: there is satisfactory agreement between the 
empirical values of $\alpha$ and the values obtained from (\ref{eq: 3.6}). 
Right panel: the data points collapse onto a 
plot of $F(p)\equiv (1-P_1)/Q$, using values of $P_k$ derived from simulations
(with $N_{\rm thr}=8$), compared with the values of $F(p)$ obtained from 
equations (\ref{eq: 5.2.1})-(\ref{eq: 5.2.5}). 
}
\end{figure}

Figure \ref{fig: 5.2} provides direct tests of
the theoretical prediction for the exponent 
$\alpha$. We estimated $\alpha$ from simulations with a wide range of values of 
$r$ and $p$. In the left panel of Fig.~\ref{fig: 5.2}, we compare 
the theoretical values of $\alpha$ 
obtained from equation (\ref{eq: 3.7}) against numerically estimates, 
obtained by directly simulating the model: here we used the values of $P_k$ obtained
from simulations with $N_{\rm thr}=8$.
In the right panel we collapse all of the data points onto two plots of 
$F(p)\equiv (1-P_1)/Q$, one obtained using equations (\ref{eq: 5.2.1}) to (\ref{eq: 5.2.5}),
the other using values of $P_k$ obtained
from simulations with $N_{\rm thr}=8$. The former shows small but significant deviations 
at larger values of $p$.  The small dispersion between the symbols gives an indication of 
the accuracy of our determination of the exponents $\alpha$.

We remark that, if the errors in the approximations underlying 
equations (\ref{eq: 5.2.1})-(\ref{eq: 5.2.5}) and (\ref{eq: 4.1.5}), (\ref{eq: 4.1.6}) 
are negligible as $p\to 0$, we can determine the limiting value of $F(p)$ as $p\to 0$. 
Using these expressions in equation (\ref{eq: 3.8}) results in
\begin{equation}
\label{eq: 5.2.6}
\lim_{p\to 0}F(p)=3
\ .
\end{equation}

\subsection{Two-dimensional model with four branches}
\label{sec: 5.3}

\begin{figure}
	\includegraphics[width=10cm]{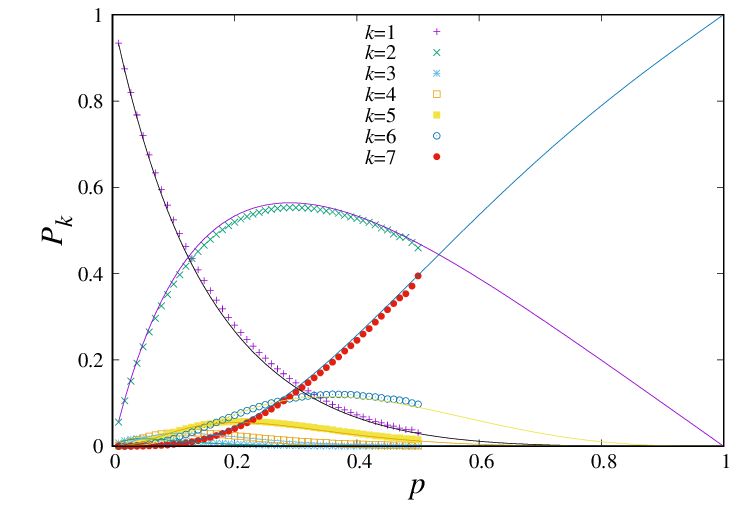}
\caption{
\label{fig: 5.3}
Numerical simulations of the functions $P_k(p)$, $k=1,\ldots,7$, for the 
two-dimensional, four-branch model.
Two different numerical simulations are shown:
one includes all sites (i.e. $N_{\rm thr}=0$, solid line), the other (points) imposes the 
requirement that the site has not experienced a coalescence in the preceding 
$N_{\rm thr}=6$  iterations.
}
\end{figure}

\begin{figure}
	\includegraphics[width=8.2cm]{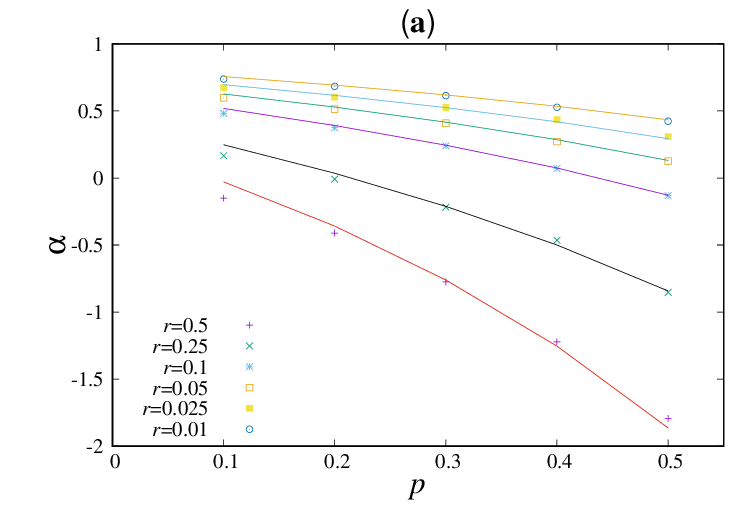}
        \includegraphics[width=8.2cm]{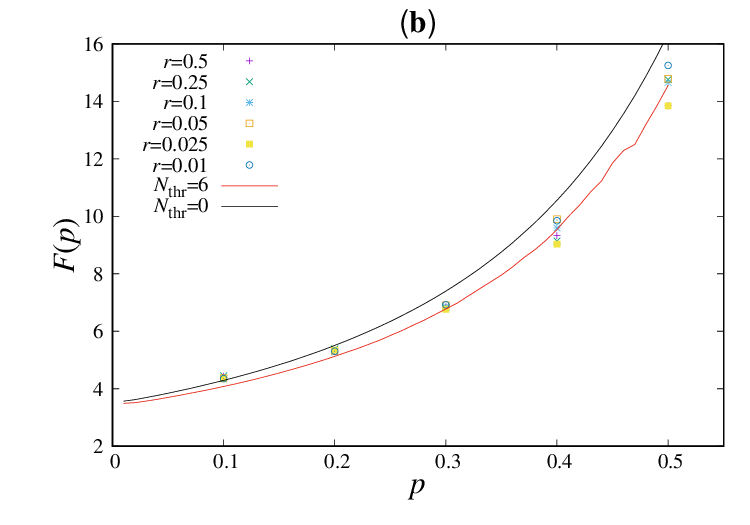}
\caption{
\label{fig: 5.4}
Testing the determination of $\alpha$ for the two-dimensional four-branch 
model, using (\ref{eq: 3.6}), and the $P_k$ derived from simulations 
with $N_{\rm thr}=6$. Left panel: there is satisfactory agreement between the 
empirical values of $\alpha$ and the values obtained from (\ref{eq: 3.6}). 
Right panel: the data points onto a plot of $F(p)\equiv (1-P_1)/Q$, using 
values of $P_k$ derived from simulations (with $N_{\rm thr}=6$), compared with 
the values of $F(p)$ obtained from simulations which included 
all sites (i.e. setting $N_{\rm thr}=0$): there is 
a significant discrepancy as $p\to 1$.
}
\end{figure}

In the case of the four-branch model, a theoretical calculation of the probabilities 
$P_k$ is considerably more difficult, although we  
can obtain formulae for $P_1$ and $P_2$ which are analogous to those 
obtained in the two-branch case, and find 
\begin{equation}
\label{eq: 5.3.1}
P_1=1-7p+O(p^2)
\end{equation}
The calculation of the other $P_k$ is complicated, 
because once a site has branched from one site to its four neighbours, one
has to consider transitions from the eight sites adjacent to the four 
newly occupied positions. Four of them can only reach one of the four 
new positions, but four of them could possibly reach two positions. We can, however, assert
that $P_3=p+O(p^2)$ and conclude that 
\begin{equation}
\label{eq: 5.3.2}
\lim_{p\to 0}F(p)=\frac{3}{2}
\ .
\end{equation}
Numerical investigation of the probabilities $P_k$ for the four-way splitting model 
is difficult because, except when $p$ is small, the proportion of sites which do not 
undergo coalescence events is very small. Accordingly, we confined our numerical 
investigations to cases where $p<0.5$. Figures \ref{fig: 5.3} and \ref{fig: 5.4}, illustrating
investigations of the four-branch model, are similar to figures \ref{fig: 5.1} and 
\ref{fig: 5.2}, but do not include theoretical predictions of $F(p)\equiv (1-P_1)/Q$.

\section{Tests of exactness in the mass-conserving case}
\label{sec: 6}

In the Introduction, we mentioned that, in one dimension, the case where 
$p_0=0$ appears to be exactly solvable. 
Here, we argue that the steady-state probability for occupying $N$ consecutive sites 
at step $j$ can be written as a product of independent probabilities
at different sites.  That is, if $s_i$ is the occupation of the 
$i^{\rm th}$ site, then we postulate the joint probability 
to be the product:
\begin{equation}
\label{eq: 6.1}
P_N(s_1,s_2,\ldots,s_N)
=\prod_{i=1}^N P_1(s_i)=\prod_{i=1}^N \left[s_i f_1+(1-s_i)(1-f_1)\right]
\end{equation}
where $f_1$ is the probability of occupation of a single site, 
given by (\ref{eq: 4.1}). This result can be demonstrated by 
assuming that (\ref{eq: 6.1}) holds at iteration $j$, and testing 
whether the joint probability given by Eq.~\ref{eq: 6.1}
remains unchanged by the dynamics. 

Because sites which are not adjacent to each other are not influenced by 
common sites at the previous iteration, non-adjacent pairs are obviously 
independent. In subsection \ref{sec: 6.1} we investigate the joint probability $P_2(a,b)$ 
for two adjacent sites, and show that this factorises if we make an assumption 
about the relationship between the occupation probability, $f$, and 
the splitting probability, $p$. This relation need not necessarily be the 
same, as the function $f_1(p)$ given by (\ref{eq: 4.1}), and we shall 
distinguish it by denoting this function by $\tilde f_1(p)$. The same 
approach is also used to determine functions $\tilde f_{2,2}(p)$ and $\tilde f_{2,4}(p)$ 
which would ensure that $P_2(a,b)=P_1(a)P_1(b)$ for the two-dimensional 
models. We show that $\tilde f_1(p)$ coincides with $f_1(p)$, implying 
that the one-dimensional model is exactly solvable, but that this does not 
hold for the two-dimensional cases.

We are also able to give an inductive demonstration of a more general result
concerning the dynamics of the one-dimensional model: in section \ref{sec: 6.2} 
we show that the boundary 
between an occupied and an unoccupied region fluctuates 
diffusively, while the occupation probabilities within the occupied 
region remain statistically independent. 

\subsection{Condition for factorisation}
\label{sec: 6.1}

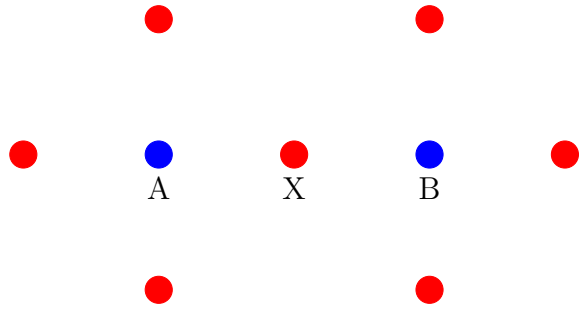
\begin{figure}
\centering
\begin{tikzpicture}[scale=0.9]
\draw[fill,red] (1,3) circle (0.2);
\draw[fill,blue] (3,3) circle (0.2);
\draw[fill,red] (3,1) circle (0.2);
\draw[fill,red] (3,5) circle (0.2);
\draw[fill,red] (5,3) circle (0.2);
\draw[fill,blue] (7,3) circle (0.2);
\draw[fill,red] (7,1) circle (0.2);
\draw[fill,red] (7,5) circle (0.2);
\draw[fill,red] (9,3) circle (0.2);
\node[] at (3,2.5) {A};
\node[] at (5,2.5) {X};
\node[] at (7,2.5) {B};
\end{tikzpicture}
\caption{
At iteration $j+1$, sites A an B are influenced by their nearest neighbours
(illustrated here for the two-dimensional model) at iteration $j$. Correlations 
may result from the fact that both A and B are influenced by occupation of 
the \lq key site', X.
}
\label{fig: 6.1}
\end{figure}

Consider two adjacent sites at step $j+1$. These 
were influenced by the configuration at step $j$ through 
their nearest neighbours. There is only one site, which 
will be referred to as the \lq key site', which can influence
both of the sites at step $j+1$ (see figure \ref{fig: 6.1}). So if we are seeking to 
establish whether sites are independent, this one 
site should receive special attention. This observation is 
also true in higher dimensions.

Consider the influence of the key site, ${\rm X}$, upon its two nearest 
neighbours, ${\rm A}$ and ${\rm B}$. 
A \lq wetted bond' connection may (probability $\tilde P_1$), or may not,
(probability $\tilde P_0$), be made from the key site ${\rm X}$ 
to site ${\rm A}$. 
Let $a=0$ or $a=1$ indicate whether site ${\rm A}$ is 
(respectively) empty or occupied. Also let $P(a\vert 0)$ be the probability that ${\rm A}$ 
becomes occupied at step $j+1$, given that no connection 
has been made to ${\rm A}$ from ${\rm X}$, 
and $P(a\vert 1)$ be the probability that  ${\rm A}$ is occupied if a connection is made 
from site ${\rm X}$. Clearly, $P(a\vert 1)=a$, because ${\rm A}$ is definitely occupied
if the connection is made, so that $P(1\vert 1)=1$ and $P(0\vert 1)=0$. With these definitions,
we can write an expression for the probability of ${\rm A}$ being occupied at 
step $j+1$. The probability $P_1(a)$ of ${\rm A}$ being occupied 
has a contribution from a term where no connection is made from ${\rm X}$ to ${\rm A}$ 
(with probability $\tilde P_0$), multiplied by the probability for the independent 
event in which site ${\rm A}$ achieves occupancy $a$ by connections from its other
neighbouring sites. Adding another term, representing events 
where ${\rm X}$ does connect to ${\rm A}$, we have
\begin{equation}
\label{eq: 6.1.1}
P_1(a)=P(a\vert 0)\tilde P_0+P(a\vert 1)\tilde P_1
\ .
\end{equation}

We shall need expressions for $\tilde P_0$ and $\tilde P_1$. These depend upon which 
version of the model we consider. 

\noindent

{\bf One-dimensional model} 

In the one-dimensional case, 
we assume that sites are occupied with probability $f$ at step $j$. So the probability of ${\rm X}$ 
being occupied and splitting is $fp$, and of ${\rm X}$ occupied and moving to the left without splitting
is $f(1-p)/2$. Hence
\begin{equation}
\label{eq: 6.1.2}
\tilde P_1=\left(\frac{1+p}{2}\right)f
\ ,\ \ \ 
\tilde P_0=1-\left(\frac{1+p}{2}\right)f
\ .
\end{equation}
Now consider the joint occupation probability of sites ${\rm A}$ and ${\rm B}$. Going from 
iteration $j$ to $j+1$, there is a probability $\tilde P_{00}$ that there is no transfer of 
occupation from ${\rm X}$ to either ${\rm A}$ or ${\rm B}$. The probability that
 ${\rm X}$ transfers to ${\rm B}$ and not ${\rm A}$ is $\tilde P_{01}$, 
 and the probability to transfer to both sites is $\tilde P_{11}$. In the one-dimensional case, these 
probabilities are
\begin{equation}
\label{eq: 6.1.3}
\tilde P_{00}=1-f
\ ,\ \ \ 
\tilde P_{01}=\frac{f}{2}(1-p)
\ ,\ \ \ 
\tilde P_{11}=pf
\ .
\end{equation}
The joint probability at step $j+1$ for sites ${\rm A}$, ${\rm B}$ is
\begin{eqnarray}
\label{eq: 6.1.4}
P_2(a,b)&=&P(a|0)P(b|0)\tilde P_{00}
\nonumber \\
&+&\left[P(a|0)P(b|1)+P(a|1)P(b|0)\right]\tilde P_{01}
\nonumber \\
&+&P(a|1)P(b|1)\tilde P_{11}
\ .
\end{eqnarray}
Now use (\ref{eq: 6.1.4}) and (\ref{eq: 6.1.1}) to determine 
\begin{eqnarray}
\label{eq: 6.1.5}
P_2(a,b)-P_1(a)P_1(b)&=&P(a|0)P(b|0)\left[\tilde P_{00}-\tilde P_0^2\right]
\nonumber \\
&+&\left[P(a|0)P(b|1)+P(a|1)P(b|0)\right]\left[\tilde P_{01}-\tilde P_0 \tilde P_1\right]
\nonumber \\
&+&P(a|1)P(b|1)\left[\tilde P_{11}-\tilde P_1^2\right]
\ .
\end{eqnarray}
From (\ref{eq: 6.1.5}), the occupations remain independent 
at step $j+1$ if all three of the following conditions are satisfied:
\begin{equation}
\label{eq: 6.1.6}
\tilde P_{00}=\tilde P_0^2
\ ,\ \ \ 
\tilde P_{10}=\tilde P_0\tilde P_1
\ ,\ \ \ 
\tilde P_{11}=\tilde P_1^2
\ .
\end{equation}
In the one-dimensional case, these conditions are
\begin{eqnarray}
\label{eq: 6.1.7}
1-f&=&\left[1-\left(\frac{1+p}{2}\right)f\right]^2
\nonumber \\
(1-p)\frac{f}{2}&=&\left[1-\left(\frac{1+p}{2}\right)f\right]\left(\frac{1+p}{2}\right)f
\nonumber \\
pf&=&\left[\left(\frac{1+p}{2}\right)f\right]^2
\end{eqnarray}
These three equations all imply the same relationship between 
$f$ and $p$:
\begin{equation}
\label{eq: 6.1.8}
\tilde f_1(p)=\frac{4p}{(1+p)^2}
\end{equation}
which is the same relationship between $f$ and $p$ as arises independently 
from the calculation of the filling fraction in the independent-site approximation, equation
(\ref{eq: 4.1}). 

Note that this calculation didn't require $P(a|0)$ 
or $P(b|1)$, only the probabilities $\tilde P_0$, $\tilde P_{00}$, $\tilde P_{01}$. 
It is, therefore, easily extended to the two-dimensional models that 
we considered. 

\noindent
{\bf Two-dimensional model}

For the two-branch model
\begin{eqnarray}
\label{eq: 6.1.9}
&&\tilde P_{00}=1-\frac{f}{2}
\ ,\ \ \ 
\tilde P_{01}=\left(\frac{1-p}{4}\right)f
\ ,\ \ \ 
\tilde P_{11}=p\frac{f}{2}
\nonumber \\
{\bf two}&-&{\bf branch\ model}
\nonumber \\
&&\tilde P_0=1-\left(\frac{1+p}{4}\right)f
\ ,\ \ \ 
\tilde P_1=\left(\frac{1+p}{4}\right)f
\ .
\end{eqnarray}
In this case, equations (\ref{eq: 6.1.7}) are satisfied by
\begin{equation}
\label{eq: 6.1.10}
\tilde f_{2,2}(p)=\frac{8p}{(1+p)^2}
\ .
\end{equation}
Similarly, for the four-branch model
\begin{eqnarray}
\label{eq: 6.1.11}
&&\tilde P_{00}=1-\left(\frac{1+p}{2}\right)f
\ ,\ \ \ 
\tilde P_{01}=\left(\frac{1-p}{4}\right)f
\ ,\ \ \ 
\tilde P_{11}=pf
\nonumber \\
{\bf four}&-&{\bf branch\ model}
\nonumber \\
&&\tilde P_0=1-\left(\frac{3p+1}{4}\right)f
\ ,\ \ \ 
\tilde P_1=\left(\frac{3p+1}{4}\right)f
\ .
\end{eqnarray}
In this case, equations (\ref{eq: 6.1.7}) are satisfied by
\begin{equation}
\label{eq: 6.1.12}
\tilde f_{2,4}(p)=\frac{16p}{(3p+1)^2}
\ .
\end{equation}
In contrast with the one-dimensional model, in the two-dimensional 
cases the functions $\tilde f(p)$, which satisfy
the factorisation condition, do not agree with the functions $f(p)$ which 
describe the occupation probability of the sites 
discussed in section \ref{sec: 4}. We conclude that the 
site occupations are not independent in the two-dimensional models.

\subsection{One-dimensional case: finite intervals}
\label{sec: 6.2}

We have shown that in the one-dimensional case, the independent-occupancy 
distribution is self-reproducing under iteration of the model in the 
mass-conserving case, $p_0=0$. We can also establish a stronger result 
on exactness of solutions of the one-dimensional case. Suppose that we know that 
at iteration $j$, the occupied region is bounded: it is known that sites $n_-$ and $n_+$ 
are occupied, and that everything to the left of $n_-$ and to the right of $n_+$ 
is empty. All the sites with a parity different from $j$ are empty. Let us assume 
that all of the sites $n$ with the same parity as $j$ satisfying $n_-<n<n_+$ are 
occupied with independently with probability $f_1$, given by (\ref{eq: 4.1}). 
What can be said about the  joint distribution of occupancy at the next ($j+1$) 
iteration? 

The end values $n_-$ and $n_+$ both change by $\pm 1$, independently. 
At each end the boundary expands with probability $(1+p)/2$  or contracts
with probability $(1-p)/2$. Let us assume that the shifts of the boundary 
have been determined, and consider the joint distribution of occupation
of the other sites, conditional upon the new positions of the ends of the occupied 
interval.

We shall argue that if, at iteration $j$, the interior sites are 
independently occupied with probability $f_1$, then, they remain independently
occupied with $f_1$ at iteration $j+1$. 
Because of the short range of influence of occupation probabilities
at the next iteration, 
we can treat the two ends independently, and consider only a short interval in the 
vicinity of one end.

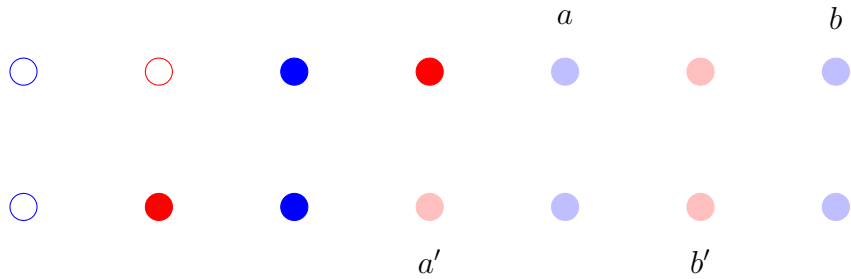
\begin{figure}
\centering
\begin{tikzpicture}[scale=0.9]
\draw[blue] (1,0) circle (0.2);
\draw[blue] (1,2) circle (0.2);
\draw[fill,red] (3,0) circle (0.2);
\draw[red] (3,2) circle (0.2);
\draw[fill,blue] (5,0) circle (0.2);
\draw[fill,blue] (5,2) circle (0.2);
\draw[fill,red!25] (7,0) circle (0.2);
\draw[fill,red] (7,2) circle (0.2);
\draw[fill,blue!25] (9,0) circle (0.2);
\draw[fill,blue!25] (9,2) circle (0.2);
\draw[fill,red!25] (11,0) circle (0.2);
\draw[fill,red!25] (11,2) circle (0.2); 
\draw[fill,blue!25] (13,0) circle (0.2);
\draw[fill,blue!25] (13,2) circle (0.2);
\node[] at (9,2.8) {$a$};
\node[] at (13,2.8) {$b$};
\node[] at (7,-0.8) {$a'$};
\node[] at (11,-0.8) {$b'$};
\end{tikzpicture}
\caption{
We consider configurations at the left-hand edge of the filled region.
At iteration $j$ (blue sites), the configuration is $(0,1,a,b)$ (where $a\in \{0,1\}$
and $b\in \{0,1\}$ are variable). At iteration $j+1$ (red sites), the occupied region either 
expands (lower row, occupancy $(1,a',b')$) or else contracts (upper row, occupancy $(0,1,b')$).
We show that if $a,b$ are independent, with occupation probability $f_1=4p/(1+p)^2$, 
then $a',b'$ have the same property. Sites defined to be empty are shown as open circles,
those defined filled are solid colour, and sites which are occupied with 
probability $\fo(p)$ are shaded.
}
\label{fig: 6.2}
\end{figure}

We consider a sequence of four sites at iteration $j$, having the same parity as $j$. 
The first two of these have definite values $0$ and $1$, and the second two are variable, 
$a\in\{0,1\}$ and $b\in\{0,1\}$, as illustrated in figure \ref{fig: 6.2}. 
These four sites, $(0,1,a,b)$, influence the values of 
sites (of different parity) at the next iteration. We consider two cases: the case where 
the occupied region \emph{expands} and the sequence at the next iteration is $(1,a',b,)$, 
and also the \emph{contracting} case where the occupations become $(0,1,b')$. 

In the contracting case, one calculates the probability of $b'$, assuming that $a$ and $b$
are independent, with probability $f_1$ of being equal to $1$:
\begin{equation}
\label{eq: 6.2.1}
{\rm Prob}(b')=\sum_{(a,b)}P_{01\vert}(b'|a,b)[f_1a+(1-f_1)(1-a)][f_1b+(1-f_1)(1-b)]
\end{equation}
where $P_{01\vert}(b'|a,b)$ is the conditional probability for obtaining $(0,1,b')$ at iteration 
$j+1$ given $(0,1,a,b)$ at iteration $j$.  Noting that the probability for a \lq contracting' 
shift of the boundary is $(1-p)/2$, if the joint distribution of occupations 
is self-reproducing under iteration, we should expect that
\begin{equation}
\label{eq: 6.2.2}
{\rm Prob}(b')=\left(\frac{1-p}{2}\right)[f_1b'+(1-f_1)(1-b')]
\ .
\end{equation}
Similarly, for the extending case, we can calculate the joint probability 
distribution of $(a',b')$:
\begin{equation}
\label{eq: 6.2.3}
{\rm Prob}(a',b')=\sum_{(a,b)}P_{1\vert}(a',b'|a,b)[f_1a+(1-f_1)(1-a)][f_1b+(1-f_1)(1-b)]
\end{equation}
where $P_{1\vert}(a',b'|a,b)$ is the conditional probability for obtaining $(1,a',b')$ at iteration 
$j+1$ given $(0,1,a,b)$ at iteration $j$. In this case, we expect
\begin{equation}
\label{eq: 6.2.4}
{\rm Prob}(a',b')=\left(\frac{1+p}{2}\right)[f_1a'+(1-f_1)(1-a')][f_1b'+(1-f_1)(1-b')]
\ .
\end{equation}
There does not appear to be any transparent general expression for the 
 conditional probabilities $P_{1\vert}(a,,b'|a,b)$ 
 and $P_{01\vert}(b'|a,b)$, and we determined
 them on a case-by-case basis. They are tabulated in table \ref{tab: 6.1}. 
 The first four rows determine $P_{1\vert}(a',b'|a,b)$, and the final two rows specify
 $P_{01\vert}(b'|a,b)$. 

\begin{table}[ht]
\caption{Conditional probabilities for reaching states $(1,a',b')$ or $(0,1,b')$ (rows)
from initial states $(0,1,a,b)$ (columns).}
\centering
\begin{tabular}{c c c c c}
\hline\hline
    &(0,1,0,0)&(0,1,0,1)&(0,1,1,0)&(0,1,1,1) \\ [0.5ex]
\hline
(1,0,0)&$\frac{1-p}{2}$&$\left(\frac{1-p}{2}\right)^2$&$0$&$0$ \\
(1,0,1)&$0$&$\left(\frac{1-p}{2}\right)\left(\frac{1+p}{2}\right)$&
$\left(\frac{1-p}{2}\right)^2$&$\left(\frac{1-p}{2}\right)^2$ \\
(1,1,0)&$p$&$\left(\frac{1-p}{2}\right)p$&
$\left(\frac{1-p}{2}\right)\left(\frac{1+p}{2}\right)$&$\left(\frac{1-p}{2}\right)^2\left(\frac{1+p}{2}\right)$ \\
(1,1,1)&$0$&$\left(\frac{1+p}{2}\right)p$&
$2p\left(\frac{1-p}{2}\right)+p^2$&
$1-\left(\frac{1-p}{2}\right)-\left(\frac{1-p}{2}\right)^2-\left(\frac{1-p}{2}\right)^2\left(\frac{1+p}{2}\right)$\\
(0,1,0)&$\left(\frac{1-p}{2}\right)$&$\left(\frac{1-p}{2}\right)^2$&
$\left(\frac{1-p}{2}\right)^2$&$\left(\frac{1-p}{2}\right)^2$\\
(0,1,1)&$0$&$\left(\frac{1-p}{2}\right)\left(\frac{1+p}{2}\right)$&
$\left(\frac{1-p}{2}\right)\left(\frac{1+p}{2}\right)$&
$\left(\frac{1-p}{2}\right)\left[1-\left(\frac{1-p}{2}\right)^2\right]$\\ [1ex]
\hline
\end{tabular}
\label{tab: 6.1}
\end{table}

Using the expressions in table \ref{tab: 6.1}, we were able to verify that 
equations (\ref{eq: 6.2.2}) and (\ref{eq: 6.2.1}) are indeed equal, as are
equations (\ref{eq: 6.2.4}) and (\ref{eq: 6.2.3}). We are not aware of any 
simpler route to this conclusion.

The same argument can be applied at the right-hand edge of the occupied region, 
so that the site occupations remain independent inside any occupied region,
independent of how its boundaries fluctuate.

\section{Relation to standard model for directed percolation}
\label{sec: 7}

Here we discuss what happens, in one dimension, when $p_0\ne 0$. 
This includes the standard model for directed percolation as a special case, and 
we shall give emphasis to making connections with that problem.
So in this section we will not consider the probability distribution of the flux $\phi$, 
but rather concentrate on the set of occupied sites. The model then has just 
two relevant parameters, $p_0$ and $p_2$.

When $p_0\ne 0$, there is a possibility 
for clusters of pathways to die out, so that the occupation fraction 
$f_1(p_0,p_2)$ is equal to zero when $p_0>p_{\rm c}$, where $p_{\rm c}$ is 
the percolation threshold. This percolation threshold is a function of $p_2$.

The standard directed bond-percolation problem is when the 
two bonds are filled independently with probability $p$, so that 
$p^2\equiv p_2$ and $(1-p)^2=p_0$. The standard
directed percolation problem is therefore represented by the 
parametric line $p_0=(1-p)^2$, $p_2=p^2$ in the two-dimensional 
parameter space of our model.

In sub-section \ref{sec: 7.1} we propose a bound on the percolation 
threshold, $p_{\rm c}(p_2)$, and compare this with numerical estimates.
Sub-section \ref{sec: 7.2} presents some numerical investigations of the 
critical exponents of the model. We find that these appear to be identical 
to those of the standard directed percolation model, in accord with 
a hypothesis of Janssen (\cite{Jan81}) and Grassberger (\cite{Gra81}). In sub-section 
\ref{sec: 7.3} we present numerical investigations of the PDF of the distribution of 
sizes of voids. 

\subsection{Phase diagram}
\label{sec: 7.1}

\begin{figure}
	\includegraphics[width=10cm]{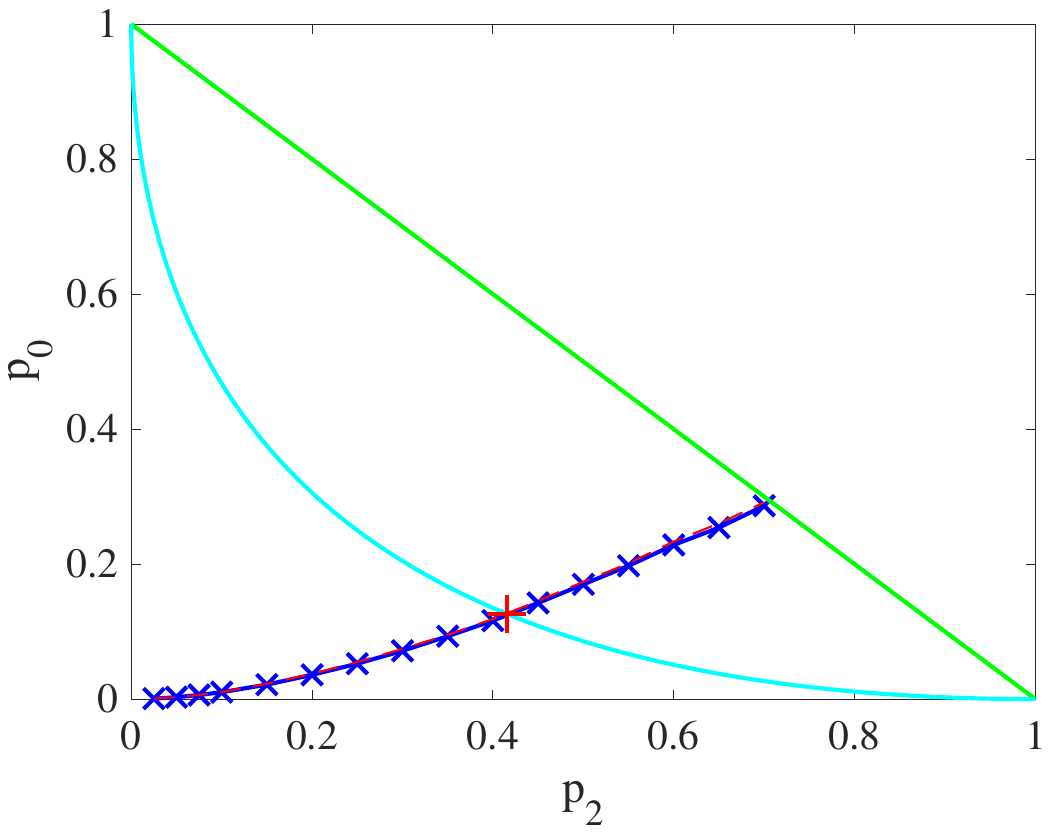}
\caption{
\label{fig: 7.1}
Parameter space of the one-dimensional model, showing the critical line 
for percolation transition (crosses). The conventional one-dimensional directed percolation 
problem, in which bonds are occupied independently with probability $p$, is 
represented parametrically by the line $p_0=(1-p)^2$, $p_2=p^2$, shown in blue.
The red cross indicates the critical point for the standard directed bond-percolation 
model, at $p\approx 0.644$.
}
\end{figure}

The space of models is illustrated in figure \ref{fig: 7.1}. The allowed region 
of $p_0$, $p_2$ space is a triangle, $p_0\ge 0$, $p_2\ge 0$, $p_0+p_2\le 1$.
The line $p_0=0$ is exactly solvable for the equilibrium distribution
as described by equation (\ref{eq: 4.1}) and section \ref{sec: 6}. 
The standard directed bond-percolation problem with probability $p$ for bond 
occupation is the line $p_0=(1-p)^2$, $p_2=p^2$. We have plotted 
our numerical evaluation of the critical line, above which 
$f_1(p_0,p_2)=0$. The critical line crosses the line defining 
bond-directed percolation at $p=0.644\ldots$, as expected \cite{Bax+88}.

We were able to suggest an upper bound on the critical line using the following 
argument (which assumes that $p_2\ll 1$).
We know that $p_0=0$ is exactly solvable, with those sites 
which are accessible on a given iteration are independently occupied 
with probability $f_1(0,p_2)= 4p_2/(1+p_2)^2$.
Because there are no correlations, the site occupation is a Poisson process. 
This expression implies that there are \lq voids' between occupied 
sites which have a characteristic lengthscale $\langle L_{\rm v}\rangle$ which 
diverges as $p_2\to 0$. Assuming that the lattice spacing is unity, and noting 
that it is only every second site which is accessible, 
the mean length of the voids is
\begin{equation}
\label{eq: 7.1.1}
\langle L_{\rm v}\rangle \sim 1/(2p_2)
\end{equation}
when $p_0=0$. These voids have 
an exponential distribution of lengths. 

A completely empty state is another possible solution. 
The dynamics of the system is 
described by the boundary between the empty state and an occupied region.
This boundary is described by a single trajectory, and its treatment 
is more tractable than analysing the joint statistics of an occupied region. 
The path of the boundary is a random walk with a drift. In the limit 
where both $p_0$ and $p_2$ are small, we can characterise the motion of the 
boundary of the occupied region by a diffusion coefficient $D_{\rm b}$ and a drift
velocity $v_{\rm b}$.

The diffusion coefficient  is determined by writing $\langle \Delta x^2\rangle=2D\Delta t$, 
and noting that the displacement is unity (except in the rare cases where the 
trajectory terminates). 
There is a drift velocity into the empty region, which is determined by noting that 
when a trajectory splits, with probability $p_2$, the daughter trajectory 
on the void side become the new boundary. When $p_0=0$ the 
diffusion coefficient $D_{\rm b}$ and a drift velocity $v_{\rm b}$
(into the empty region) are. 
\begin{equation}
\label{eq: 7.1.2}
D_{\rm b}=\frac{1}{2}
\ ,\ \ \ 
v_{\rm b}=p_2
\ .
\end{equation}
Note that after coarse-graining the spatial and temporal scales, the edge of a void 
satisfies a stochastic differential equation of the form 
\begin{equation}
\label{eq: 7.1.3}
{\rm d}x=v_{\rm b}{\rm d}t +\sqrt{2D_{\rm b}}{\rm d}\eta (t)
\end{equation}
where ${\rm d}\eta$ is a standard stochastic increment, satisfying 
$\langle {\rm d}\eta(t)\rangle=0$ and $\langle {\rm d}\eta(t){\rm d}\eta(t')\rangle =\delta (t-t')$.
This is equivalent to a Fokker-Planck equation for the position of a void boundary:
\begin{equation}
\label{eq: 7.1.4}
\frac{\partial P}{\partial t}=-\frac{\partial}{\partial x}\left[v_{\rm b}P\right]
+D_{\rm b}\frac{\partial^2 P}{\partial x^2}
\ .
\end{equation}
A similar expression holds for the overall width of the void. Taking account 
of the fact that the void has two edges, both the drift velocity and the 
diffusion coefficient are doubled.   
The steady-state probability density for the distribution of large void 
sizes is then
\begin{equation}
\label{eq: 7.1.5}
P_{\rm v}(L)={\rm const.}\times \exp(-L/\langle L_{\rm v}\rangle)
\end{equation}
where the steady state size scale for the large voids is 
$\langle L_{\rm v}\rangle =2D_{\rm b}/2v_{\rm b}\sim 1/2p_2$,
which is in agreement with equation (\ref{eq: 7.1.1}).

\begin{figure}
	\includegraphics[width=10cm]{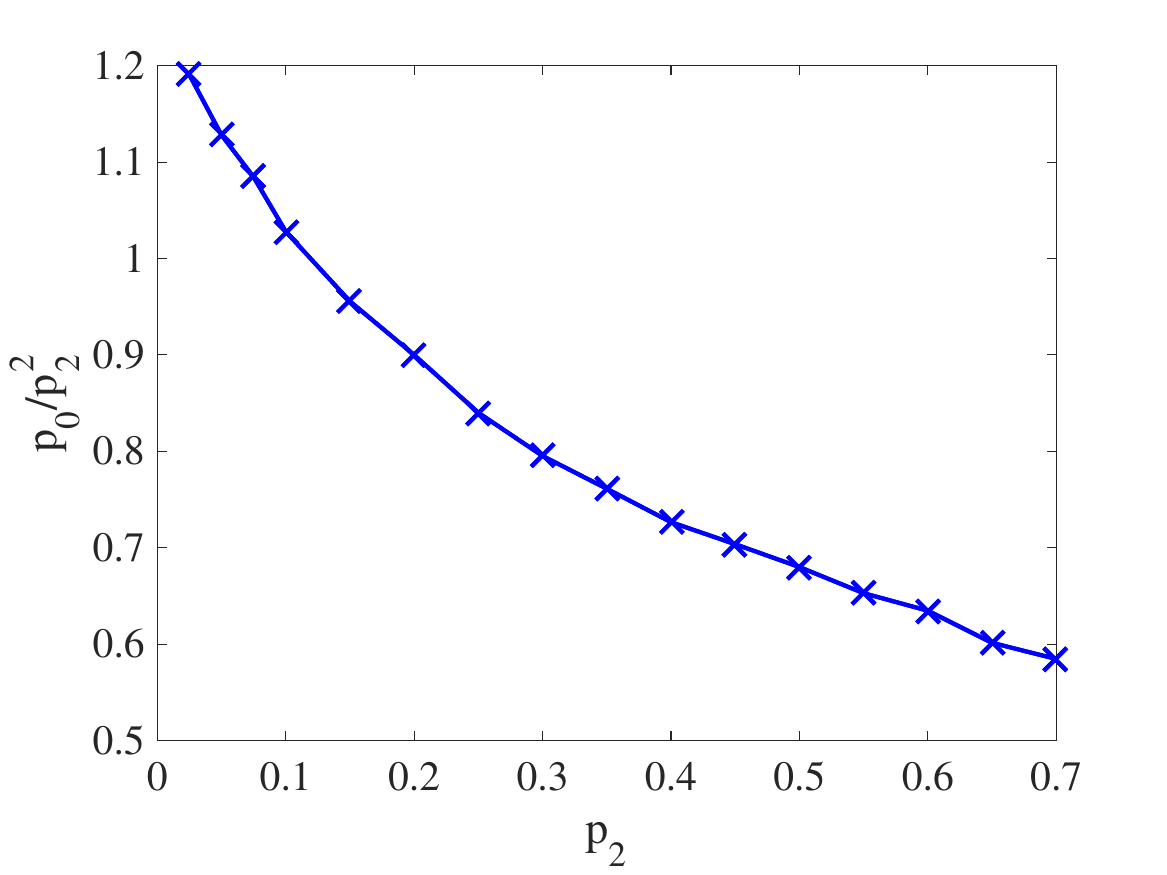}
\caption{
\label{fig: 7.2}
Ratio $p_0/p_2^2$ for transition line, showing consistency with equation 
(\ref{eq: 7.1.8}). 
}
\end{figure}

Consider what happens as $p_0$ is increased. When $p_0\ne 0$ 
there is a new mechanism which contributes to the drift velocity of 
the boundary. When a trail disappears, the boundary of the occupied 
region retreats by a distance $L_1$ equal to the length of the first void 
which is encountered. The velocity of the boundary is then 
\begin{equation}
\label{eq: 7.1.6}
v_{\rm b}=p_2-p_0\langle L_1\rangle
\end{equation}
which becomes negative at a critical value of $p_0$. 
The significance of this critical value is that for $p_0>p_{\rm c}$ the boundary 
of the occupied region retreats, until we are left with the empty configuration.
We should expect that $\langle L_1\rangle$ increases when $p_0>0$, so that
(when $p_2\ll 1$), $v_{\rm b}$ is bounded above: 
\begin{equation}
\label{eq: 7.1.7}
v_{\rm b}\le p_2-\frac{p_0}{2p_2}
\ .
\end{equation}
The critical value of $p_0$ occurs when $v_{\rm b}$ changes sign, so that
\begin{equation}
\label{eq: 7.1.8}
p_{\rm c}\le 2p_2^2
\ .
\end{equation}
In figure \ref{fig: 7.2} we plot $p_0/p_2^2$ for the transition line, as a function of 
$p_2$. The ratio is less than $2$, as predicted by equation (\ref{eq: 7.1.8}). 
We do not have a theory for the form of this dependence.

Finally, we remark that our model is related to a model of directed 
percolation in which bonds are deleted with probability $1-p_{\rm bond}$, and 
sites are deleted with probability $1-p_{\rm site}$, as discussed in \cite{Tre+95}. 
The parameters $p_{\rm bond}$ and $p_{\rm site}$ are related to our parameters 
as follows:
\begin{equation}
\label{eq: 7.1.9}
p_0=(1-p_{\rm site})+p_{\rm site}(1-p_{\rm bond})^2
\ ,\ \ \ 
p_2=p_{\rm site}p_{\rm bond}^2
\ .
\end{equation}
This bond/site deletion model does not cover the whole
parameter space of our model: the square $0\le p_{\rm site}\le 1$, 
$0\le P_{\rm bond}\le 1$ maps to the region to the right of the 
cyan line in figure \ref{fig: 7.1}. It can be verified that the region 
of the critical line of our model lying within this region is in agreement
with the data in \cite{Tre+95}.

\subsection{Critical exponents}
\label{sec: 7.2}

\begin{figure}
	\includegraphics[width=8.2cm]{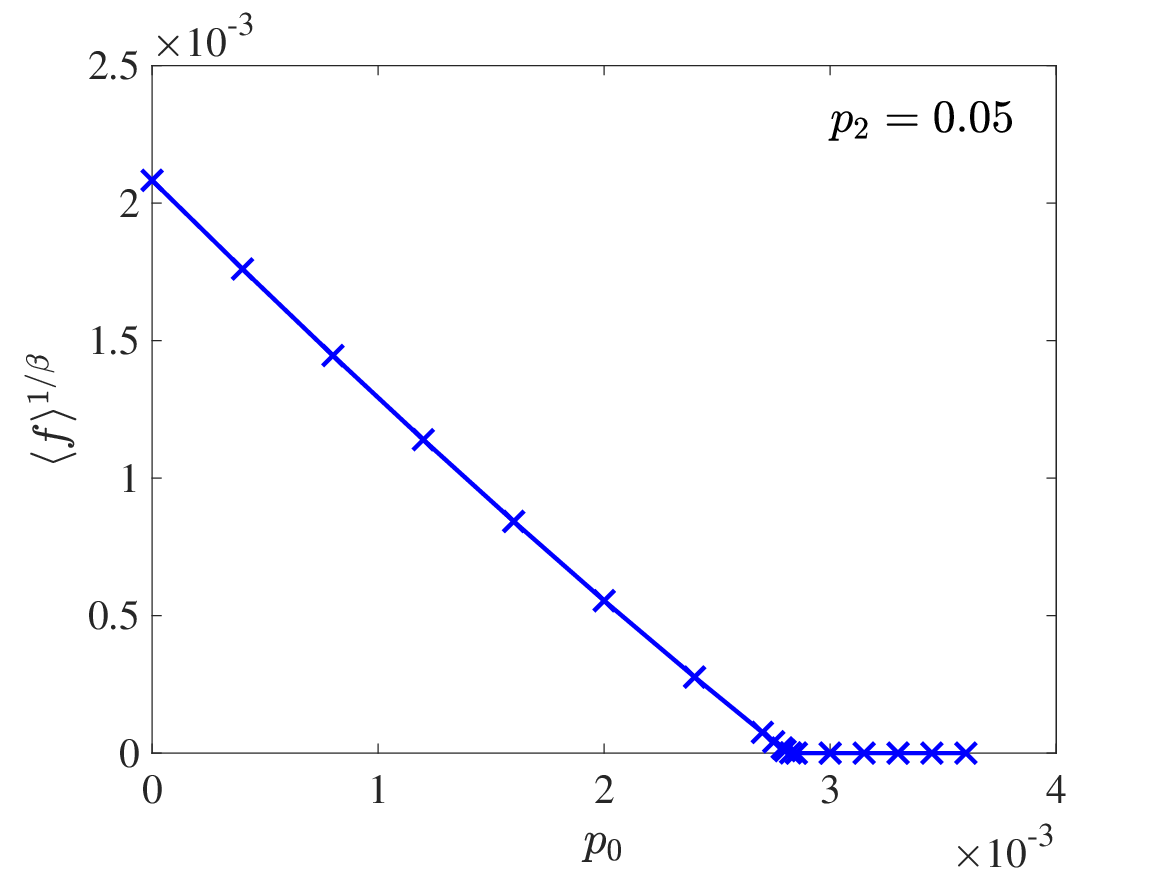}
	\includegraphics[width=8.2cm]{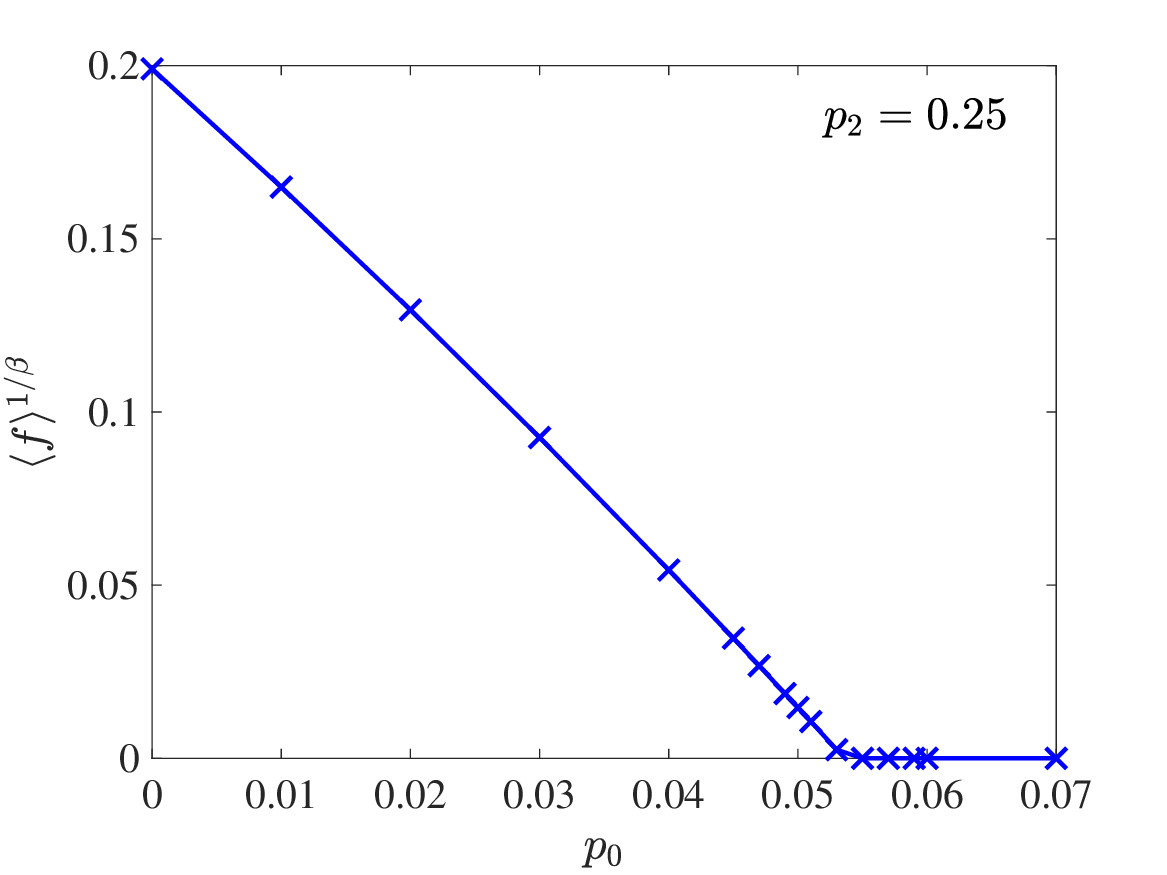}
\caption{
\label{fig: 7.3}
Occupied fraction approaching the percolation transition. 
The average fraction, to the power $1/\beta$, where $\beta$ 
is the exponent of directed percolation, shown as a function of $p_0$, reveals
an approximately linear law. This is consistent with the expected behaviour
$\langle f \rangle \propto (p_c(p_2) - p_0)^\beta$ when 
$p_0 \rightarrow p_c(p_2)$, see Eq.~(\ref{eq: 7.2.1}). 
}
\end{figure}

The percolation process which we consider is a form of directed 
percolation. Grassberger \cite{Gra81} and Janssen \cite{Jan81} 
introduced a hypothesis that the 
directed percolation has universal critical exponents, and we expect that the transition  
in our model lies in this universality class. Thus we expect that $f_1(p_0,p_2)=0$ 
when $p_0>p_{\rm c}(p_2)$, and that when $p_{\rm c}-p_0$ is small and positive we 
have 
\begin{equation}
\label{eq: 7.2.1}
f_1(p_0,p_2)\sim |p_{\rm c}(p_2)-p_0|^{\beta}
\ .
\end{equation}
where $\beta=0.276\ldots$ is the critical exponent for the order parameter 
of the directed percolation transition. Figure \ref{fig: 7.3} presents evidence that 
the occupation fraction vanishes in accord with equation (\ref{eq: 7.2.1}).

\subsection{Void size distribution}
\label{sec: 7.3}

\begin{figure}
	\includegraphics[width=8.2cm]{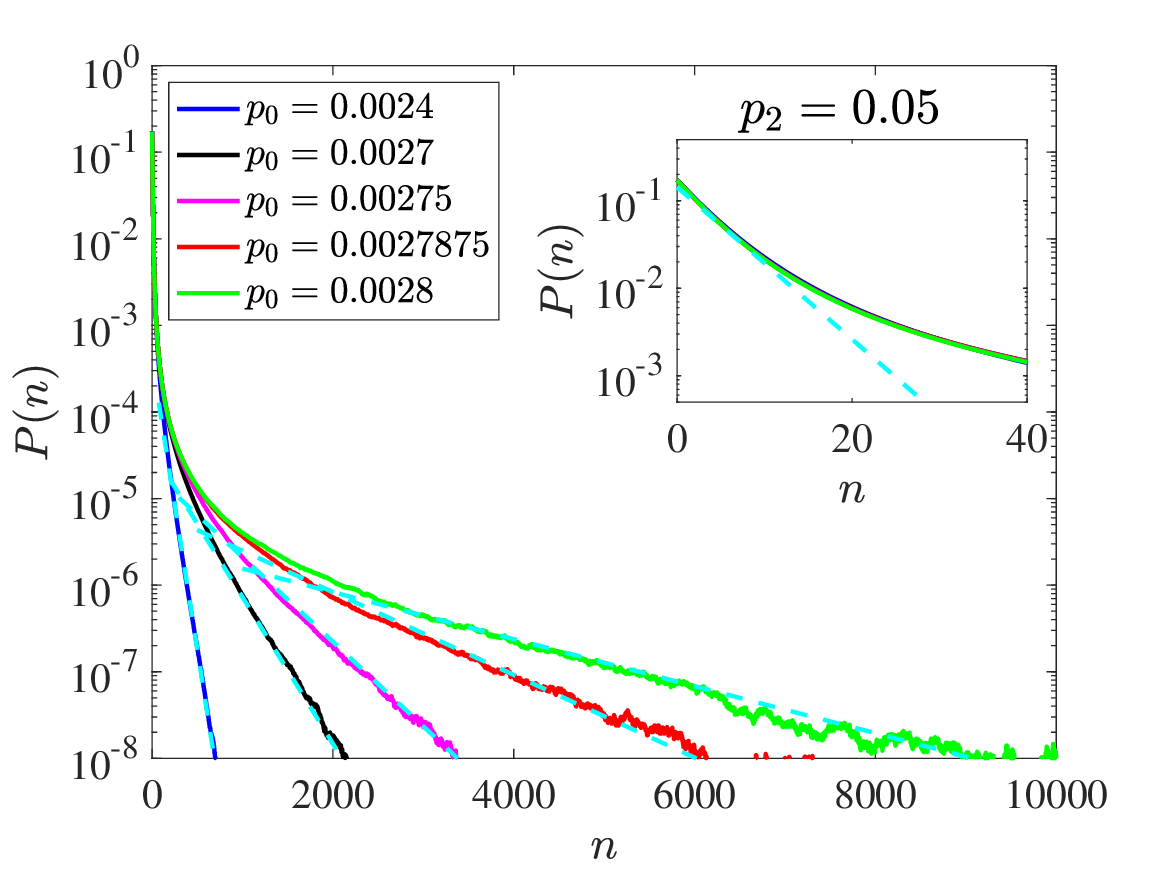}
	\includegraphics[width=8.2cm]{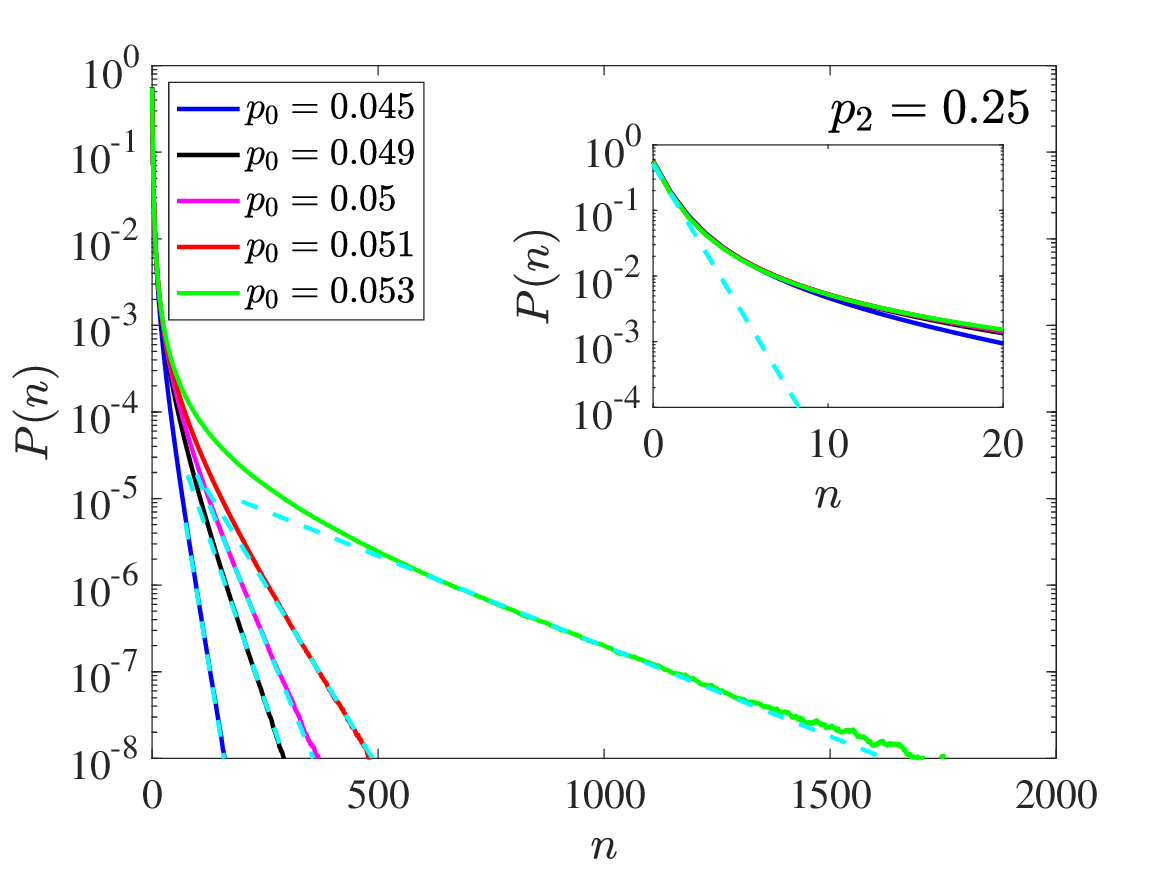}
\caption{
\label{fig: 7.4}
Void size distribution. The PDFs of the size distributions for $p_2 = 0.05$ (left) 
and $p_2 = 0.25$ (right) show broad tails, which can be approximately
represented by an exponential form: $P(n) \propto \exp( - \mu n)$; as
indicated by the dashed line in the figure. The 
exponent $\mu$ becoming very small when $p_0$ approaches the transition 
point. For small values of $n$, on the other hand, the distributions of $n$
are also exponential, consistent with equation (\ref{eq: 7.1.5}): 
$P(n) \approx \fo^n$, with $\fo = 4 p_2 /(1 + p_2^2)$,
as indicated by the dashed line in the inset. 
}
\end{figure}

We investigated the distribution of sizes of voids, $P_{\rm void}(n)$. 
The distribution, illustrated in figure \ref{fig: 7.4}  is highly distinctive: the is a \lq core' region, in which 
$P_{\rm void}(n)$ has a rapid exponential decay, and a \lq tail', which has a 
slower exponential decay, described by an exponent $\mu$: 
\begin{equation}
\label{eq: 7.3.1} 
P(n) \propto \exp( - \mu n)
\end{equation}
The lengthscale associated with the slow decay diverges 
as the critical point is approached. (In figure \ref{fig: 7.4} the void size $n$ 
is the number of \emph{potentially} filled sites which are actually empty. Because sites 
with different parity from the iteration index are automatically empty, the void sizes 
disccused in sub-section \ref{sec: 7.1} are approximately $2n$.)

\begin{figure}
	\includegraphics[width=10cm]{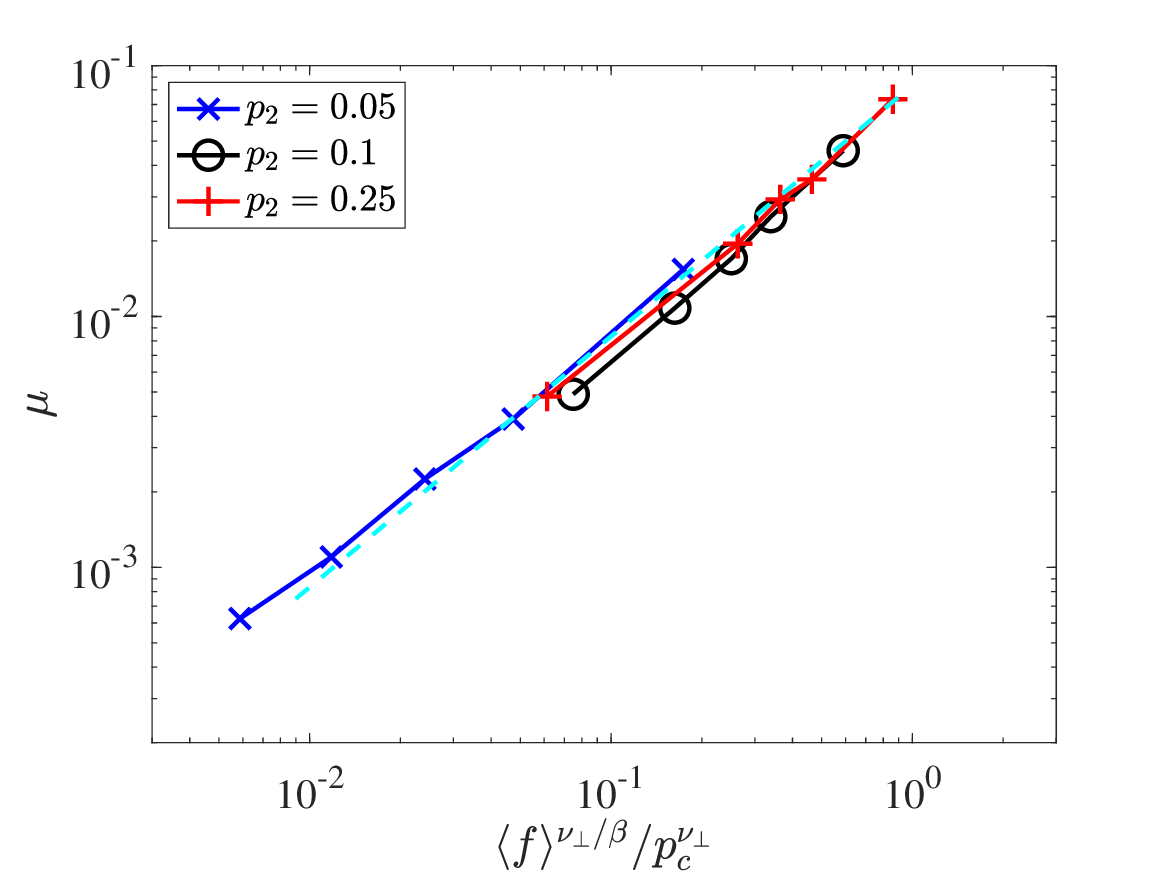}
\caption{
\label{fig: 7.5}
The decay rate $\mu$ of the void size distributions as a function of 
the \emph{normalised} mean occupation fraction, 
$\langle f \rangle/p_2$, to the power $\nu_\perp/\beta$, where $\nu_\perp$
and $\beta$ are the classical directed percolation exponents~\cite{Hin00}. The
dashed line indicate a power $1$ corresponds to the directed
percolation scaling, and describes approximately the numerical observations.
}
\end{figure}

Figure \ref{fig: 7.5} shows how values of the decay rate $\mu$ 
behaves as a function of the probability $p_0$, at the values of $p_2 = 0.05$, $0.1$ and 
$0.25$. In the spirit of directed percolation, one expects that the 
exponent $\mu \propto |p_c(p_2) - p_0|^{\nu_\perp}$ when $p_0 < p_c(p_2)$.
Since $\langle f \rangle \propto | p_c(p_2) - p_0|^{\beta}$, Fig.~\ref{fig: 7.5}
shows $\mu$ as a function of $\langle f \rangle^{\nu_\perp/\beta}$. 
Additionally, 
we divided the value of 
$\langle f \rangle^{\nu_\perp/\beta}$ by $p_c^{\nu_\perp}$, which reduces
the various values to a unique curve.
The dashed line shows
a power law with a power $1$ (linear dependence), which closely agrees with 
the measured values of $\mu$ collapsed to a single curve.
 
\section{Conclusions, implications for elution}
\label{sec: 8}

We introduced an alternative model for directed percolation which comes closer 
to addressing the questions originally posed by Broadbent and Hammersley \cite{Bro+57}. 
Our model considers the distribution of fluxes $\phi$ through wetted bonds, and 
it has a flux-conserving regime, in which the total flux remains constant. 
Our model has similarities to the Scheidegger river model \cite{Sch67}, which 
also involves addition of fluxes upon combining paths, and which also leads 
to a power-law distribution of fluxes \cite{Hub91}. The models differ as to the source 
of the flux: in our model the flux total flux is constant when $p_0=0$, whereas in the 
Scheidegger model new sources are added continuously.

We find that the distribution of fluxes $\phi$ is very broad, and in section \ref{sec: 3} we
showed that it has a power-law asymptote 
at small fluxes: $P(\phi)\sim \phi^{-\alpha}$. The exponent of this power law, $\alpha$, was 
found to vary continuously as a function of the parameters of the model. 
We found an exact equation for $\alpha$ (equation (\ref{eq: 3.6})) 
in terms of some probabilities $P_k$ 
(defined by (\ref{eq: 3.3})) which 
characterise the \lq wetting' of the skeleton of bonds through which the fluxes flow.

In one dimension, the sites are independently occupied in the steady-state, 
implying that many quantities of interest, including the exponent $\alpha$, and 
the filling fraction for occupied sites, $f_1$, can be determined exactly.
 
In two dimensions, the independent-occupancy assumption 
gives a very good approximation for $\alpha$ and for the filling fractions, as 
shown in sections \ref{sec: 4} and \ref{sec: 5}, but it is not exact.
In section \ref{sec: 6} we examined the conditions for the site occupations 
to remain independent under iteration. We found that these are only consistent
with the predicted values of the filling fraction $f$ in the one-dimensional case.

In the case where the paths can terminate, the model is no longer flux conserving, 
and we find that the distribution of wetted bonds has a percolation transition. 
In section \ref{sec: 7} we showed that, in the one-dimensional case, 
the critical exponents are in agreement with those of the standard directed percolation model. 
We also investigated the distribution of void sizes.

The long-tailed distribution of fluxes is expected to have practical consequences.
Consider the following model for an elution process. For definiteness we discuss 
a solute being washed out of a solid substrate by a liquid. Leaching of a salt 
from a permeable rock or from land reclaimed from the sea would be a concrete
example. We assume that the solid medium is permeated by randomly arranged 
narrow channels, but that on a large scale it appears homogeneous. We shall assume
that the fluid is being forced through the medium, by gravity, or a pressure difference
(or both).

If the solid contains solute with a concentration $c$ (defined as 
mass per unit volume), this will come into equilibrium with the 
salt dissolved in the liquid phase at a concentration $Kc$, where $K$ is a partition coefficient. 
If the liquid is flowing in very narrow channels, we can assume that the solute 
equilibrates between the solid and liquid phases, so that the concentration in the 
fluid flowing through the pore is also $Kc$. The rate at which solute is removed 
from the medium by flow through the pore is therefore $\dot m=Kc\phi$, where $\phi$ 
is the volume flux through the pore. The concentration is proportional to the amount 
of solute remaining in the solid phase, so that we expect that the concentration 
in the runoff through the pore reduces exponentially as a function of time. We therefore
expect that the rate of loss of solute from a single pore is
\begin{equation}
\label{eq: 8.1}
\dot m=Kc(0)\phi\exp[-\nu \phi t]
\end{equation}
where $\nu$ is dependent upon the geometry of the 
pore and the coefficient of solubility in the liquid and solid phases.

In the case of perfusion through a random medium, the liquid may 
follow many different channels (labelled by an index $i$), with very different 
volume fluxes $\phi_i$ in different channels. In this case, the total rate of 
solute out of the medium is 
\begin{equation}
\label{eq: 8.2}
\dot M(t)=\sum _i \dot m_i(t)=Kc(0)\sum_i \phi_i \exp(-\nu_i \phi_i t)
\ .
\end{equation}
We shall argue that the predominant factor determining the rate of elution 
at long times is the existence of pores which carry very low fluxes. For this reason
we adopt the simplifying assumption that the coefficients $\nu_i$ are the 
same for all of the pores. 
If the probability density function of $\phi$ is $P(\phi)$, and the density of pores 
carrying fluid is $\rho$, then the time-dependence of the 
eluted flux from a surface of area $A$ is 
\begin{eqnarray}
\label{eq: 8.3}
\dot M(t)&=&K\rho A c(0) \int_0^\infty {\rm d}\phi\ P(\phi)\phi\exp(-\nu \phi t)
\nonumber \\
&=&-K\rho A c(0)\frac{{\rm d}\bar P}{{\rm d}s}\bigg\vert_{\nu t}
\end{eqnarray}
where $\bar P(s)$ is the Laplace transform of $P(\phi)$.
Determining the long-time behaviour depends upon the distribution of $\phi$ in the limit as 
$\phi\to 0$. We have argued that this has a power-law behaviour for a wide range 
of models. This indicates that 
there are many channels which have an extremely small flux, corresponding to 
a slow elution of the solute. Using (\ref{eq: 8.3}), this corresponds 
to a power-law decay of the eluted solute flux:
\begin{eqnarray}
\label{eq: 8.4}
\dot M(t)&\sim& c(0)A\rho  \int_0^\infty {\rm d}\phi \ \phi^{1-\alpha} \exp(-\nu \phi t)
\nonumber \\
&=&\Gamma(2-\alpha) c(0)A \rho (\nu t)^{\alpha-2}
\ .
\end{eqnarray}
It is noteworthy that, due to the power-law distribution of the flux $\phi$, despite the 
fact that the elution from each channel decreases exponentially as a function of time, 
the overall rate of elution has a much slower, power-law, decay. This is a consequence 
of the log-time behaviour being dominated by the channels with the smallest flux.

{\bf Acknowledgements}. The authors are grateful to 
Robert Ziff for pointing out reference \cite{Tre+95}, and 
its relevance to our system, as expressed by equation (\ref{eq: 7.1.9}).
The project was initiated at the Kavli Institute for Theoretical Physics for support, where
this research was  supported in part by the National Science Foundation
under Grant No. PHY11-25915.

\end{document}